\documentclass[a4paper,11pt]{article}
\usepackage{jheppub} % for details on the use of the package, please
\usepackage{amsmath,amssymb,amsthm,amscd,graphicx,xcolor}
\usepackage{slashed}
\usepackage[subrefformat=parens,labelformat=parens]{subcaption}
\input epsf.sty

\usepackage{texdraw}
\theoremstyle{definition}

\newcommand{\CB}{{\cal B}}

\newcommand{\CE}{{\cal E}}
\newcommand{\CF}{{\cal F}}

\newcommand{\CI}{{\cal I}}

\newcommand{\CL}{{\cal L}}

\newcommand{\CO}{{\cal O}}
\newcommand{\CP}{{\cal P}}

\newcommand{\CS}{{\cal S}}

\newcommand{\CX}{{\cal X}}
\newcommand{\CY}{{\cal Y}}
\newcommand{\CZ}{{\cal Z}}

\def\IZ{{\mathbb Z}}

\newcommand{\bS}{\boldsymbol{S}}

\newcommand{\mQ}{\mathsf{Q}}

\newcommand{\mH}{\mathsf{H}}

\newcommand{\bPsi}{\boldsymbol{\Psi}}

\newcommand{\re}{{\rm e}}
\newcommand{\ri}{{\rm i}}
\newcommand{\rd}{{\rm d}}

\newcommand{\be}{\begin{equation}}
\newcommand{\ee}{\end{equation}}
\newcommand{\ba}{\begin{aligned}}
\newcommand{\ea}{\end{aligned}}
\newcommand{\ben}{\begin{eqnarray}\displaystyle}
\newcommand{\een}{\end{eqnarray}}

\newcommand{\bpsi}{\boldsymbol{\psi}}
\newcommand{\sectiono}[1]{\section{#1}\setcounter{equation}{0}}

\DeclareMathOperator{\disc}{disc}
\DeclareMathOperator{\Res}{Res}
\DeclareMathOperator{\Tr}{Tr}
\let\div\relax
\DeclareMathOperator{\div}{div}
\let\Im\relax
\DeclareMathOperator{\Im}{Im}

\makeatletter
\g@addto@macro\bfseries{\boldmath}
\makeatother

\let\originalleft\left
\let\originalright\right
\renewcommand{\left}{\mathopen{}\mathclose\bgroup\originalleft}
\renewcommand{\right}{\aftergroup\egroup\originalright}

\newcommand{\widebar}[1]{\mkern 1.5mu\overline{\mkern-1.5mu#1\mkern-1.5mu}\mkern 1.5mu}

\newdimen\tableauside\tableauside=1.0ex
\newdimen\tableaurule\tableaurule=0.4pt
\newdimen\tableaustep
\def\phantomhrule#1{\hbox{\vbox to0pt{\hrule height\tableaurule width#1\vss}}}
\def\phantomvrule#1{\vbox{\hbox to0pt{\vrule width\tableaurule height#1\hss}}}
\def\sqr{\vbox{%
  \phantomhrule\tableaustep
  \hbox{\phantomvrule\tableaustep\kern\tableaustep\phantomvrule\tableaustep}%
  \hbox{\vbox{\phantomhrule\tableauside}\kern-\tableaurule}}}
\def\squares#1{\hbox{\count0=#1\noindent\loop\sqr
  \advance\count0 by-1 \ifnum\count0>0\repeat}}
\def\tableau#1{\vcenter{\offinterlineskip
  \tableaustep=\tableauside\advance\tableaustep by-\tableaurule
  \kern\normallineskip\hbox
    {\kern\normallineskip\vbox
      {\gettableau#1 0 }%
     \kern\normallineskip\kern\tableaurule}%
  \kern\normallineskip\kern\tableaurule}}
\def\gettableau#1{\ifnum#1=0\let\next=\null\else
\squares{#1}\let\next=\gettableau\fi\next}

\newcommand{\figref}[1]{figure~\protect\ref{#1}}
\usepackage{mathtools}

\title{\Huge{\boldmath Testing the Bethe ansatz with large $N$ renormalons}}

\author{Marcos Mari\~no,}
\author{Ramon Miravitllas}
\author{and Tom\'as Reis}

\affiliation{D\'epartement de Physique Th\'eorique et Section de Math\'ematiques\\
Universit\'e de Gen\`eve, Gen\`eve, CH-1211 Switzerland}

\emailAdd{Marcos.Marino@unige.ch}
\emailAdd{Ramon.MiravitllasMas@unige.ch} 
\emailAdd{Tomas.Reis@unige.ch} 

\abstract{
The ground state energy of integrable asymptotically free theories can be conjecturally computed by using the Bethe ansatz, once the theory has been coupled to an external potential through a conserved charge. This leads to a precise prediction for the 
perturbative expansion of the energy. We provide a non-trivial test of this prediction in the non-linear sigma model and its supersymmetric extension, by calculating analytically the associated Feynman diagrams at next-to-leading order in the $1/N$ expansion, and at all loops. By investigating the large order behaviour of the diagrams, we locate the position of the renormalons of the theory and we obtain an analytic expression for the large $N$ trans-series associated to each. As a spin-off of our calculation, we provide a direct derivation of the beta function of these theories, at next-to-leading order in the $1/N$ expansion.}

\begin{document}
\maketitle
\flushbottom
 
\section{Introduction}
 
In quantum field theory, the conventional perturbative series explores just the fluctuations around the trivial vacuum. 
It is believed that a full picture should involve additional sectors, corresponding to non-trivial vacua, and leading to exponentially suppressed 
contributions to physical observables. The origin of 
these sectors might be additional semi-classical saddles, also called instantons, but in 
many cases, like in Yang--Mills theory, the most important non-perturbative effects are associated to elusive objects 
called renormalons \cite{beneke}.  

 The contribution of instanton sectors can be obtained in principle by expanding the path integral around these non-trivial saddle-points. 
 In the case of renormalons, we do not have such a semi-classical picture. It is known however that renormalons manifest themselves in the large order behavior of perturbation theory \cite{parisi1,parisi2}. Therefore, one way 
to obtain precise quantitative information about renormalon sectors and their exponentially small corrections 
is to use conventional perturbation theory at large number of loops. This connection, in which a non-perturbative correction ``resurges" in the 
perturbative series, is part of the general theory of resurgence, which gives a systematic framework to understand non-perturbative sectors (see \cite{mmlargen,mmbook,ss,abs,du-review} for introductions). 

In spite of some impressive lattice calculations \cite{pineda} (see also \cite{pcf-lattice}), unveiling the resurgent structure of 
realistic quantum field theories remains a difficult problem. One could however 
try to address these issues in more tractable quantum field theories. Among these, 
integrable field theories in two dimensions play an important r\^ole. 
On the one hand, they display many of the physical phenomena of interest, like asymptotic freedom 
and the presence of renormalon sectors. On the other hand, they are much more tractable analytically. 
In particular, exact expressions for their $S$ matrices have been conjectured in \cite{zz, zamo-zamo}.

It was noted by Polyakov and Wiegmann in \cite{pw,wiegmann2} that, in integrable field theories, 
one can calculate exactly the dependence of the ground state 
energy on a chemical potential coupled to a global conserved charge. This is done by combining the 
exact $S$-matrix with the Bethe ansatz, and the 
answer can be elegantly expressed in terms of a set of integral equations. 
One beautiful application of this observation was the determination of the exact mass gap 
of these theories, by comparing the Bethe ansatz to conventional 
perturbation theory (see \cite{hmn,hn, fnw1, fnw2,eh-ssm} for various case 
studies, and \cite{eh-review} for a review). The ground state energy as a function of the chemical potential, 
which we will call for short the free energy of the theory, is then an ideal observable to understand 
quantitatively the different sectors, perturbative and non-perturbative, of the theory. 

However, even extracting the perturbative expansion of the free energy from the 
Bethe ansatz answer remained a challenge for a long time (this is a generic difficulty of the Bethe ansatz, 
which goes back to the Lieb--Liniger solution of the interacting Bose gas in one dimension \cite{ll}). 
An algorithm to do this was finally found by 
Volin in \cite{volin, volin-thesis}. This opened the way to a resurgent analysis 
of quantum integrable systems. The presence of renormalons in the non-linear sigma model was tested 
numerically in \cite{volin}. In \cite{mr-ren}, 
this analysis was extended to many two-dimensional asymptotically free theories, and some classical predictions of renormalon asymptotics 
were verified in detail. A similar resurgent study of various non-relativistic models was carried out in \cite{mr-long, mr-ll, mr-hub}. 
Very precise results on the resurgent structure of the free energy in the $O(4)$ sigma model 
have been recently obtained in \cite{abbh1, abbh2}, where Volin's method was used to generate the perturbative series up to very large order.

In this paper we study the free energy of the $O(N)$ non-linear sigma model 
and its supersymmetric extension. Our first goal is a detailed test of the Bethe ansatz result against conventional 
perturbation theory. So far, this has been verified up to two-loops in \cite{bbbkp}. In order 
to do a comprehensive test, we calculate the free energy to \emph{all loops}, at next-to-leading order in the $1/N$ expansion. 
Our result matches the Bethe ansatz answer obtained in \cite{volin,mr-ren} to all available orders. In this way, it 
provides very convincing evidence for the validity of the Bethe ansatz, as well as for the techniques of \cite{volin,mr-ren} to extract perturbative series from it. It also provides additional evidence for the conjectural $S$ matrices (although these have been tested directly, up to next-to-leading order in the $1/N$ expansion, in \cite{berg, gracey}). We should note that, in order 
to make contact with perturbation theory, our $1/N$ expansion is performed around the perturbative vacuum, 
and not around the non-trivial large $N$ vacuum where particles get a non-perturbative 
mass already at tree level. In this sense, our calculation is very similar to what was done in \cite{mr-new}: we re-organize perturbation theory in powers of $1/N$ and keep the first two non-trivial contributions.

An interesting spin-off of our calculation is a new derivation 
of the beta function of the non-linear sigma model, at next-to-leading order in the $1/N$ expansion. This is a known result, going back to \cite{bh}. 
However, in \cite{bh} the beta function was derived from the epsilon expansion of the critical exponents 
(see \cite{gracey-rev} for a review). Our derivation is a direct one, similar in spirit and in the details to the 
computation of Palanques-Mestre and Pascual of the QED beta function in the limit of large number of fermions 
$N_f$ \cite{pmp}. In the case of the supersymmetric non-linear sigma model, 
we verify the vanishing of the beta function
at next-to-leading order in $1/N$, which was also derived from the epsilon expansion of critical exponents in 
\cite{gracey-super}. 

In some cases, one can calculate the $1/N$ expansion of the free energy directly from the Bethe ansatz 
equations \cite{fnw2,fkw1,fkw2, zarembo, ksz,dmss}. However, even if the 
Bethe ansatz result contains complete information about the ground state energy, including both the 
perturbative expansion and non-perturbative corrections, at present no known method exists to 
extract analytically the non-perturbative sectors, not even 
at large $N$. In contrast, our calculation of the perturbative series at next-to-leading order in $1/N$ is fully analytic. By using
 techniques developed in \cite{mr-new,mr-hub, mr-roads}, we can extract the trans-series at large $N$ 
 from our all-loop results. This trans-series signals the presence of an 
 isolated infrared (IR) renormalon singularity\footnote{In \cite{volin, volin-thesis} it was erroneously concluded, from the behavior of the very first terms of the perturbative series, that there are no IR renormalons at this order in $1/N$.}, and an 
 infinite sequence of ultraviolet (UV) renormalon singularities. 
 
In the supersymmetric extension of the non-linear sigma model, an important 
consequence of our perturbative computation is an analytic proof, at next-to-leading order in $1/N$, that the IR 
singularity arising from bosonic diagrams does not cancel against the IR singularity arising from fermionic diagrams. Therefore, the 
supersymmetric model exhibits IR renormalons\footnote{In the first version of \cite{mr-ren}, the numerical results for the Borel--Pad\'e transform of the perturbative series of this model were interpreted as an indication that the first IR renormalon is absent. This is incorrect. A more careful analysis of this series confirms the presence of this IR renormalon, also at finite $N$.}. This is in contrast to the disappearance of leading IR renormalons in some supersymmetric theories, pointed out \cite{shifman-susy,dsu}. 

The organization of this paper is as follows. In section \ref{rev-sec} we review basic 
aspects of the non-linear sigma model, its $1/N$ expansion, 
and its Bethe ansatz solution. Section \ref{effpot-sec} presents the calculation of the 
effective potential at next-to-leading order in $1/N$. This result is then used in 
section \ref{ren-sec} to extract the large $N$ renormalons and their trans-series. In \ref{susy-sec} we extend all these results to the 
supersymmetric non-linear sigma model. Finally, in \ref{conclude-sec} we present 
our conclusions and prospects for future developments. There are three Appendices 
with additional details and clarifications on our calculations.

\sectiono{The non-linear sigma model and its Bethe ansatz solution}
\label{rev-sec}
The $O(N)$ non-linear sigma model is a quantum field theory in two dimensions for a 
vector field $\bS(x)= (S_1(x), \dots, S_N(x))$, satisfying the constraint
\be
\label{cons}
\bS^2=1. 
\ee
The Lagrangian density is
\be
\CL={1\over 2 g_0^2}  \partial_\mu {\boldsymbol{S}} \cdot  \partial^\mu {\boldsymbol{S}}, 
\ee
where $g_0$ is the bare coupling constant. The non-linear sigma model is asymptotically free \cite{polyakov}, and it can be regarded as a toy model for gauge theories. It also has many different applications in condensed matter physics, where it is used to model the low-energy dynamics of 
statistical systems with a global $O(N)$ symmetry. We will write the beta function for the coupling constant $g$ as 
\be
\label{betaf}
\beta_g (g)= \mu {\rd g \over \rd \mu} = -\beta_0 g^3 - \beta_1 g^5- \cdots, 
\ee
%
%and we will denote 
%
%\be
%\label{xi-constant}
%\xi = \frac{\beta_1}{2 \beta_0^2}.
%\ee
%
With this convention, asymptotically free theories have $\beta_0>0$. 
All of our perturbative calculations will be done in the ${\overline{\text{MS}}}$ scheme. For the non-linear sigma model, 
the first two coefficients of the beta function are \cite{bzj} 
\be
\beta_0= {1 \over 4 \pi \Delta}, \qquad \beta_1= {1\over 8 \pi^2 \Delta}, 
\ee
where
\be
\label{Delta2} \Delta= {1\over N-2}.  
\ee
The beta function is known up to four loops \cite{bh, wegner}. 

The non-linear sigma model can be also studied in the limit in which the number of components of $\bS$ is large and relevant quantities are computed in a large $N$ expansion. In this setting, one introduces the 't Hooft coupling 
\be
\lambda_0 = {g_0^2 \over 2 \pi \Delta},
\ee
which is kept fixed in the large $N$ limit. As we will see, it will be more natural to make the expansion in powers of $\Delta$, rather than in $1/N$.

The large $N$ expansion makes it possible to obtain non-perturbative results for this model (see e.g. \cite{mmbook}). 
In perturbation theory, one expands around an ordered vacuum with $\langle \bS \rangle \not=0$, in which the global $O(N)$ symmetry is spontaneously broken. This leads to a perturbative spectrum consisting of $N-1$ Goldstone bosons. This can not be the 
case at the non-perturbative level, due to the Coleman--Mermin--Wagner theorem. Indeed, at large $N$, 
one finds a spectrum consisting of $N$ massive particles in the fundamental representation of $O(N)$, which is thought to be the 
true spectrum of the theory. It is important however to keep in mind that perturbation theory 
around the ``false vacuum" gives the correct asymptotic expansion of $O(N)$ invariant observables, as pointed out in \cite{jevicki,elitzur}. 

Many quantities in the non-linear sigma model can be calculated systematically in a $1/N$ expansion, like for example critical exponents. 
By using this method, one finds that the beta function for the 't Hooft coupling, defined as 
\be
\beta(\lambda)=\mu {\partial \lambda \over \partial \mu},
\ee
has the following $1/N$ expansion \cite{bh}:
\be
\label{betaN}
\beta(\lambda) = \sum_{\ell\ge 0} \beta_{(\ell)}(\lambda) \Delta^\ell,
\ee
%\ee
%
where  
\begin{align}
\beta_{(0)}(\lambda)&=-\epsilon \lambda-\lambda^2, \label{beta-bh-0}\\
%\beta_1(\lambda) &= -\lambda^2 \int_0^{\lambda}  \rd x \,  \frac{2}{\pi}\left(\frac{1+x}{2+x}\right)\frac{\sin\left(\frac{\pi x}{2}\right)}{\pi x}\frac{\Gamma(1+x)}{\Gamma\left(1+\frac{x}{2}\right)^2},\\
\beta_{(1)}(\lambda) &= -4\lambda^2 \int_0^{\lambda}  \rd x\, \frac{\sin(\frac{\pi x}{2})}{\pi x}\frac{\Gamma(1+x)}{\Gamma(1+\frac{x}{2})^2}  \frac{x+1}{x+2} , \label{beta-bh-1}
\end{align}
and $\epsilon$ is related to the number of dimensions of the theory by 
\be
\label{defeps}
d=2-\epsilon.
\ee
As a spin-off of the results in this paper, we will rederive the result for $\beta_{(1)}(\lambda)$ by a direct calculation in perturbation theory.

A conjectural expression for the exact $S$-matrix of the two-dimensional non-linear sigma model was put forward in \cite{zz,zamo-zamo}. This makes it possible the following exact computation \cite{pw}. 
Let $\mH$ be the Hamiltonian of the model, and let $\mQ$ be a conserved charge, associated to a global conserved 
current. Let $h$ be an external field coupled to $\mQ$. $h$ can be regarded as a chemical potential, and as usual in statistical mechanics we can consider the ensemble defined by the operator
\be
\label{HQ}
\mH- h \mQ.
\ee
The corresponding free energy per unit volume is then defined by 
\be
\label{free-en}
\CF(h) =-\lim_{V, \beta \rightarrow \infty} {1\over V\beta } \log \Tr \re^{-\beta (\mH-h \mQ)}, 
\ee
where $V$ is the volume of space and $\beta$ is the total length of Euclidean time. This is the ground state energy of the model in the presence of the additional coupling. 
As pointed out in \cite{pw}, we can compute 
\be
\label{cfh}
\delta \CF (h)= \CF(h)- \CF(0) 
\ee
by using the exact $S$ matrix and the Bethe ansatz. We will refer to (\ref{cfh}) as the free energy. 
After turning on the chemical potential $h$ beyond an appropriate 
threshold, there will be a density  $\rho$ of particles, charged under $\mQ$, and with an energy per unit volume given by $e(\rho)$. These two quantities can 
be obtained from the density of Bethe roots $\chi(\theta)$. This density is supported on an interval $[-B,B]$ and satisfies the integral equation
\be
\label{chi-ie}
m \cosh \theta=\chi(\theta)-\int_{-B}^B  \rd \theta' \, K(\theta-\theta') \chi (\theta').
\ee
In this equation, $m$ is the mass of the charged particles, and with a clever choice of $\mQ$, it is directly related to the mass gap of the theory. 
The kernel of the integral equation is given by 
\be
K(\theta)={1\over 2 \pi \ri} {\rd \over \rd\theta} \log S(\theta),
\ee
where $S(\theta)$ is the $S$-matrix appropriate for the scattering of the charged particles. The energy per unit volume and the density are then given by
\be
\label{erho}
e={m \over 2 \pi} \int_{-B}^B \rd \theta \, \chi(\theta)  \cosh \theta, \qquad \rho={1\over 2 \pi} \int_{-B}^B \rd \theta\, \chi(\theta). 
\ee
Let us note that $B$ is fixed by the value of $\rho$, and this leads implicitly to a function $e(\rho)$. 
Finally, the free energy can be obtained as a Legendre transform of $e(\rho)$:
\be
\label{legendre}
\ba
\rho&=-\delta \CF'(h), \\
\delta \CF(h)&=e(\rho)-\rho h.
\ea
\ee
Note that the first equation defines $\rho$ as a function of $h$. 

The above program can be implemented in a number of models. In the case of the non-linear sigma model, 
one considers the conserved currents associated to the 
global $O(N)$ symmetry, 
\be
J_{\mu}^{ij}=S^{i}\partial_{\mu}S^{j}-S^{j}\partial_{\mu}S^{i}.   
\ee
We will denote by $Q^{ij}$ the corresponding charges. Usually \cite{hmn, hn}, 
one considers in \eqref{HQ} the quantum version of $Q^{12}$. The exact $S$ matrix of the $O(N)$ non-linear sigma model, 
for particles charged under $\mQ^{12}$, is given by
\be
S(\theta)=-{\Gamma (1+\ri x) \Gamma({1\over 2} +\Delta +\ri x) \Gamma({1\over 2}-\ri x) \Gamma(\Delta-\ri x) \over 
\Gamma (1-\ri x)\Gamma({1\over 2}+\Delta-\ri x) \Gamma({1\over 2}+\ri x) \Gamma(\Delta +\ri x) }, \qquad x={\theta \over 2 \pi}
\ee
and $\Delta$ is given in \eqref{Delta2}.

The perturbative series can be extracted from the Bethe ansatz solution with a method developed by Volin \cite{volin, volin-thesis}.
In his original work, the method was applied to the non-linear sigma model, but it was later extended to other quantum integrable models in \cite{mr-ren,mr-long, mr-ll,mr-hub}. It is convenient to use an expansion variable which can be connected directly to the perturbative answer. Such a variable was introduced in \cite{bbbkp} and is defined by\footnote{A scheme for QCD closely related to (\ref{a-tilde}) was studied in \cite{jamin-mira}.}
\be\label{a-tilde}
\frac{1}{\alpha}+(\xi-1)\log \alpha = \log\left(\frac{\rho}{2\beta_0 \Lambda} \right), 
\ee
where 
\be
\label{xi-def}
\xi=\frac{\beta_1}{2\beta_0^2}=\Delta
\ee
and $\Lambda$ is the dynamically generated scale
\be
\label{Lambda}
\Lambda=\mu \left( 2 \beta_0 g^2 \right)^{-\beta_1/(2 \beta_0^2)} \re^{-1/(2 \beta_0 g^2)} 
\exp\left( -\int_0^g \left\{ {1\over \beta_g (x)}+ {1\over \beta_0 x^3} -{\beta_1 \over \beta_0^2 x} \right\} \rd x  \right).
\ee
This scale is proportional to the mass $m$ appearing in the Bethe ansatz. In the case of the non-linear sigma model, one has \cite{hmn,hn}:
\be
m =\left( {8 \over \re} \right)^{\Delta} {1\over \Gamma(1+ \Delta)} \Lambda. 
\ee
 In this way, we obtain a power series in $\alpha$ for the normalized energy density, 
\be
\label{ealpha}
{e(\rho) \over \rho^{2}\pi \Delta}= \alpha \sum_{n \ge 0} a_n \alpha^n. 
\ee
 The result, up to order $\alpha^4$, is 
 \begin{multline}
\label{eperta4}
{e(\rho) \over \rho^{2}\pi \Delta} = \alpha +\frac{\alpha ^2}{2}+\frac{\alpha ^3 \Delta }{2}\\
+\frac{\Delta}{32} \alpha ^4  \left(-8 \Delta ^2 (3 \zeta (3)+1)+14 \Delta  (3 \zeta (3)+2)-21 \zeta (3)+8\right)+ \CO(\alpha^5). 
\end{multline}

The free energy $\delta\CF(h)$ can also be calculated in perturbation theory from the effective potential. The coupling to the conserved charge $Q^{12}$ leads to the modified Lagrangian \cite{hmn,hn,bbbkp}
\be
\label{hLag}
\CL_h = \frac{1}{2 g_0^2}\bigg\{ \partial_\mu \bS\cdot \partial^\mu\bS+2\ri h (S_1\partial_0 S_2-S_2 \partial_0 S_1)+h^2\left(S_3^2+\cdots+S_N^2-1\right)\bigg\}.
\end{equation}
$\delta \CF(h)$ was obtained at one-loop in \cite{hmn, hn}, and at two-loops in \cite{bbbkp}. The result of the perturbative calculation can be expressed in various convenient ways, depending on an appropriate choice of coupling. We can use the renormalization group (RG) to re-express the perturbative series in terms of the coupling $\bar g^2 (\mu/h, g)$, defined by 
\be
\label{RG-evol}
\log\left( {\mu \over h} \right)= -\int_{g}^{{\bar g}} {\rd x \over \beta_g(x)}, 
\ee
or in terms of the dynamically generated scale defined in \eqref{Lambda}. In terms of $\bar g$, the free energy is 
\be
\label{fginv}
\delta \CF(h) = -\frac{h^{2}}{2} \left\{ \frac{1}{\bar{g}^{2}}
-\beta_0-{\beta_1\over 2} {\bar g}^{2}+\CO\left(\bar g^4\right)\right\}.
\ee
In order to compare with the result of the Bethe ansatz, we can use the Legendre transform of 
$\delta \CF (h)$ to calculate $e(\rho)$, and then re-express the result in terms of the coupling $\alpha$ introduced in \eqref{a-tilde}. We note that, 
at leading order in the coupling expansion, 
\be
\bar g^2= {1\over 2\beta_0} \alpha + \CO(\alpha^3). 
\ee
One obtains in this way from perturbation theory, and up to two-loops \cite{bbbkp}, 
\begin{equation} \label{epsrho}
{e(\rho) \over \rho^{2}\pi \Delta}= \alpha+ \frac{\alpha^2}{2}
+\frac{\alpha^{3}\Delta}{2}+\CO\left(\alpha^{4}\right).
\end{equation}
This is in agreement with the result of the Bethe ansatz calculation (\ref{eperta4}), as verified in \cite{volin,bbbkp}. 

It is convenient to organize the free energy as a $1/N$ expansion: 
\be
\label{deltaN}
\delta \CF(h) = \sum_{\ell \ge 0} \delta \CF_{(\ell)}(h)  \, \Delta^{\ell-1}. 
\ee
Similarly, the normalized energy density in (\ref{ealpha}) can be organized as a double power series expansion in $\alpha$, $\Delta$:
\be
\label{ela}
{e(\rho) \over \rho^{2}\pi \Delta} = \sum_{\ell \ge 0} \CE_{(\ell)} (\alpha) \Delta^\ell, 
\ee
where $\CE_\ell (\alpha)$ are power series in $\alpha$. One finds, at leading order, 
\be
\label{E0-alpha}
\CE_{(0)}(\alpha)= \alpha+ \frac{\alpha^2}{2}, 
\ee
and at subleading order in $\Delta$ we have, for the very first terms \cite{volin,mr-ren}, 
\be
\label{e1-ba}
\CE_{(1)} (\alpha)= \frac{\alpha^{3}}{2}+ \left(\frac{1}{4}-\frac{21 \zeta (3)}{32}\right) \alpha^4 + \left(\frac{1}{4} +\frac{35 \zeta
   (3)}{32}\right) \alpha^5+ \left(\frac{3}{8}-\frac{735 \zeta (3)}{512}-\frac{4185 \zeta
   (5)}{2048}\right) \alpha^6+ \CO\left(\alpha^7\right). 
\ee
The series $\CE_{(1)} (\alpha)$ has been computed analytically up to order $44$ in \cite{mr-ren}. In the 
next section we will compute $\CE_{(1)}(\alpha)$ analytically and at all loops, directly in perturbation theory, and we will match the 
result \eqref{e1-ba} and up to order 44.

\sectiono{The \texorpdfstring{$1/N$}{1/N} expansion of the effective potential} \label{sec_effective_potential}

\label{effpot-sec}

To evaluate $\delta \mathcal{F}(h)$ in perturbation theory we have to calculate the effective potential in the theory with Lagrangian \eqref{hLag}. 
As in other problems with an $O(N)$ symmetry, it is convenient to reformulate the model in terms of 
a linear sigma model, by including an additional field $X$ which implements the constraint \eqref{cons}. In this way we consider
\begin{equation}
\CL_h =
\frac{1}{2g_0^2}\bigg\{ \partial_\mu \boldsymbol{S}\cdot \partial^\mu\boldsymbol{S}
+2\ri h \left(S_1\partial_0 S_2-S_2 \partial_0 S_1\right)+h^2(S_3^2 + \dots + S_N^2 - 1) + X(\boldsymbol S^2 - 1) \bigg\}.
\end{equation}
We expand around the following classical vacuum:
\begin{equation}
\ba
\boldsymbol\sigma(x) &= \Big(\sigma,0,\dots,0 \Big) + \sqrt{2\pi\Delta\lambda_0}\Big(\tilde\sigma_1(x),\tilde\sigma_2(x),\eta_1(x),\dots,\eta_{N-2}(x) \Big),\\
X(x) &= \chi +  \sqrt{2 \pi \Delta \lambda_0} \tilde{\chi}(x),
\ea
\end{equation}
where $\sigma$ and $\chi$ are constants that minimize the potential. Neglecting linear terms, the resulting Lagrangian can be organized as
\begin{equation}
\label{lag-nlsmh}
\CL_h = \frac{1}{2\pi\Delta\lambda_0}\CL_\text{tree} + \CL_\text{G} + \sqrt{2\pi\Delta\lambda_0}\CL_\text{int}.
\end{equation}
The tree-level Lagrangian is
\begin{equation}
\CL_\text{tree} = \frac{\chi}{2}(\sigma^2-1) - \frac{h^2}{2},
\end{equation}
while the quadratic or Gaussian part is given by
\begin{equation}
\CL_\text{G} = \frac{1}{2}\boldsymbol\eta \cdot (-\partial^2 + \chi + h^2) \boldsymbol\eta + \frac{1}{2}
\left[
\begin{matrix}
\tilde\sigma_1\\
\tilde\sigma_2\\
\tilde\chi
\end{matrix}
\right]^T
\left(
\begin{matrix}
-\partial^2 + \chi & 2ih\partial_0 & \sigma\\
-2ih\partial_0 & -\partial^2 + \chi & 0\\
\sigma & 0 & 0
\end{matrix}
\right)
\left[
\begin{matrix}
\tilde\sigma_1\\
\tilde\sigma_2\\
\tilde\chi
\end{matrix}
\right],
\end{equation}
with $\boldsymbol\eta = (\eta_1,\dots,\eta_{N-2})$. The interaction part is
\begin{equation}
\CL_\text{int} = \frac{1}{2}\tilde\chi \boldsymbol\eta \cdot \boldsymbol\eta + \frac{1}{2}\tilde\chi \Big( \tilde\sigma_1^2 + \tilde\sigma_2^2 \Big).
\end{equation}
In the theory with $h=0$, there is an ordered phase with $\chi=0$, $\sigma\not=0$ in which $\tilde \sigma_2$, $\boldsymbol{\eta}$ are 
Goldstone bosons. This phase is not realized quantum-mechanically, as we explained in the previous section. However, once $h$ is turned on, 
these $\boldsymbol{\eta}$ bosons acquire a mass.

\begin{figure}
\centering
\includegraphics{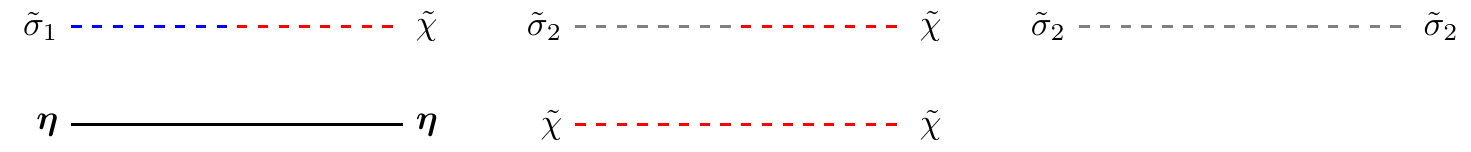}
\caption{Propagators arising in the non-linear sigma model when its Lagrangian is expanded around the classical vacuum.}
\label{nlsmprops}
\end{figure}

After writing the quadratic terms $\CL_\text{G}$ in momentum space and inverting the matrix for the fields $(\tilde{\sigma}_1,\tilde{\sigma}_2,\tilde{\chi})$, we obtain the propagators
\begin{equation}
\ba
D_{\tilde{\sigma}_1\tilde{\sigma}_1} &=  D_{\tilde{\sigma}_1\tilde{\sigma}_2} = D_{\tilde{\sigma}_2\tilde{\sigma}_1}= 0, &D_{\tilde{\sigma}_2\tilde{\sigma}_2} &= \frac{1}{k^2+\chi},\\
D_{\tilde{\sigma}_1\tilde{\chi}} &= D_{\tilde{\chi}\tilde{\sigma}_1} =\frac{1}{\sigma}, &D_{\tilde{\sigma}_2\tilde{\chi}} & = - D_{\tilde{\chi}\tilde{\sigma}_2} = \frac{2 h k_0}{\sigma(k^2+\chi)}, \\
D_{\tilde{\chi}\tilde{\chi}}&= -\frac{1}{\sigma^2}\left[k^2+\chi+ \frac{4 h^2 k_0^2}{k^2+\chi}\right], &
D_{\eta_i \eta_j} &= \delta_{ij} \frac{1}{k^2+h^2+\chi}, \quad i,j=1, \dots, N-2. 
\ea
\label{nlsmLkin}
\end{equation}
They are represented by the lines in \figref{nlsmprops}.

\begin{figure}
\centering
\includegraphics{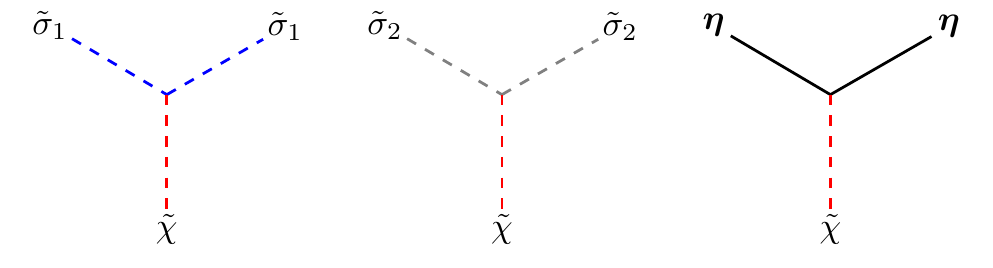}
\caption{Interaction terms arising in the non-linear sigma model when its Lagrangian is expanded around the classical vacuum.}
\label{nlsmvert}
\end{figure}

From the interaction terms in $\CL_\text{int}$, we have three cubic vertices coupling the 
field $\tilde \chi $ with $\boldsymbol{\eta}$ and $\tilde \sigma_i$, $i=1$, $2$. They are represented in \figref{nlsmvert}. 

\subsection{Effective potential at one loop}
The effective potential is a function of the vaccum expectation values (vev) $\sigma$ and $\chi$, and the parameter $h$. It has a $1/N$ expansion in powers of $\Delta$ given by
\be
\label{v-delta}
V(\sigma, \chi; h) = \sum_{\ell \ge 0} V_{(\ell)}(\sigma, \chi; h) \Delta^{\ell-1}. 
\ee
The vevs $\sigma$ and $\chi$ are obtained by extremizing the potential, and they can be also expanded in $1/N$:
\be
\sigma= \sigma_{(0)}+ \CO\left(\Delta\right), \qquad \chi= \chi_{(0)}+ \CO\left(\Delta\right). 
\ee
The leading order vevs $\sigma_{(0)}$ and $\chi_{(0)}$, which are the only ones needed in our calculation, are obtained as 
\be
{\partial V_{(0)}\over \partial \sigma_{(0)}}= {\partial V_{(0)}\over \partial \chi_{(0)}}=0.
\label{v-minima}
\ee
At next-to-leading order in $\Delta$, we have 
\be
\begin{aligned}
\delta \CF(h)= \frac{1}{\Delta} V_{(0)}(\sigma_{(0)}, \chi_{(0)};h)+ V_{(1)}(\sigma_{(0)}, \chi_{(0)};h)+ \CO\left(\Delta\right).
\end{aligned}
\ee

Let us first compute $V_{(0)}(\sigma, \chi;h)$. It has contributions from the tree level Lagrangian $\CL_\text{tree}$ and from the $1/\Delta = N-2$ fields $\eta$ at one-loop:
\begin{equation}
V_{(0)} (\sigma,\chi;h) =  \frac{1}{4 \pi  \lambda_0 } \left(\chi(\sigma^2-1)-h^2 \right)+\frac{1}{2} \int \frac{\rd^d k}{(2\pi)^d}\log(k^2+h^2+\chi).
\end{equation}
By using dimensional regularization to evaluate the integral, we find
\be
V_{(0)}(\sigma, \chi;h)= {1 \over 4 \pi \lambda_0} \left( \chi (\sigma^2-1)- h^2 \right)+ {(h^2+ \chi)^{d/2} \over  (4 \pi)^{d/2}} {1 \over d} \Gamma\left({\epsilon \over 2} \right).
\label{v-leading-order}
\ee
From here we can calculate $\sigma_{(0)}$ and $\chi_{(0)}$. There is a ``disordered" non-perturbative vacuum with $\chi_{(0)}\not=0$ and an ``ordered" perturbative vacuum with $\chi_{(0)}=0$. As in \cite{mr-new}, in order to make contact with conventional perturbation theory, we choose the perturbative vacuum, where $\chi_{(0)}=0$. Imposing \eqref{v-minima}, we find
\be
\label{sigma0}
{\sigma^2_{(0)} \over \lambda_0} = {1\over \lambda_0}- \frac{1}{2} \left(\frac{h^2}{4\pi}\right)^{-\epsilon} \Gamma\left({\epsilon \over 2} \right).
\ee
%
%where we have denoted
%
%\be
%H^2= {h^2 \over 4 \pi}. 
%\ee
%
We can now introduce the renormalized coupling $\lambda$ by the usual equation, 
\be
\lambda_0= \nu^{\epsilon} Z \lambda, 
\ee
where 
\be
\nu^2= \mu^2 \re^{\gamma-\log(4\pi)}
\ee
parametrizes the scale choice $\mu$ and features the $\overline{\text{MS}}$ scheme, and $\gamma$ is the Euler–Mascheroni constant. $Z$ is the renormalization constant, for which we consider the $1/N$ expansion
%and $Z^{-1}$ has a $1/N$ expansion of the form 
%
\be
\label{largeZ}
Z^{-1} = \sum_{\ell \ge 0} Z_{(\ell)}^{-1} \Delta^\ell. 
\ee
Cancelation of singular terms in the r.h.s. of \eqref{sigma0} fixes
\be
\label{Z0}
Z_{(0)}^{-1}=  1 +{ \lambda \over  \epsilon}. 
\ee
%
%In terms of the renormalized coupling we obtain 
%
%\be
%\label{sigma-0}
%{\lambda_0 \over\sigma_{(0)}^2}= {\nu^{\epsilon} \lambda \over 1+  \lambda \left( {1 \over \epsilon}- {1\over 2} \left( {H \over \nu} \right)^{-\epsilon} %\Gamma({\epsilon\over 2}) \right)} + \CO(\Delta). 
%\ee
%
Now, the r.h.s. of \eqref{sigma0} is manifestly finite as $\epsilon \rightarrow 0$, and we find 
\be
{\sigma_{(0)}^2 \over \lambda_0 }= {1\over \lambda} +  \log\left({h \over \mu} \right)+ \CO\left(\Delta, \epsilon\right).
\label{sigma^2/lambda_limit}
\ee
By evaluating the leading order effective potential \eqref{v-leading-order} at the critical point $\sigma_{(0)}$, $\chi_{(0)}=0$, we obtain the leading order free energy $\delta \CF_{(0)}(h)$, defined in \eqref{deltaN}. After writing the resulting expression in terms of the renormalized coupling $\lambda$ and in the limit $\epsilon \rightarrow 0$, one finds
\be
\delta \CF_{(0)}(h)=-{h^2 \over 4 \pi} \left\{ {1 \over \lambda}  + \log\left( {h \over \mu} \right) - \frac{1}{2}  \right\}. 
\ee

\subsection{Ring diagrams}
\label{ring-diagrams}

\begin{figure}
\centering
\includegraphics{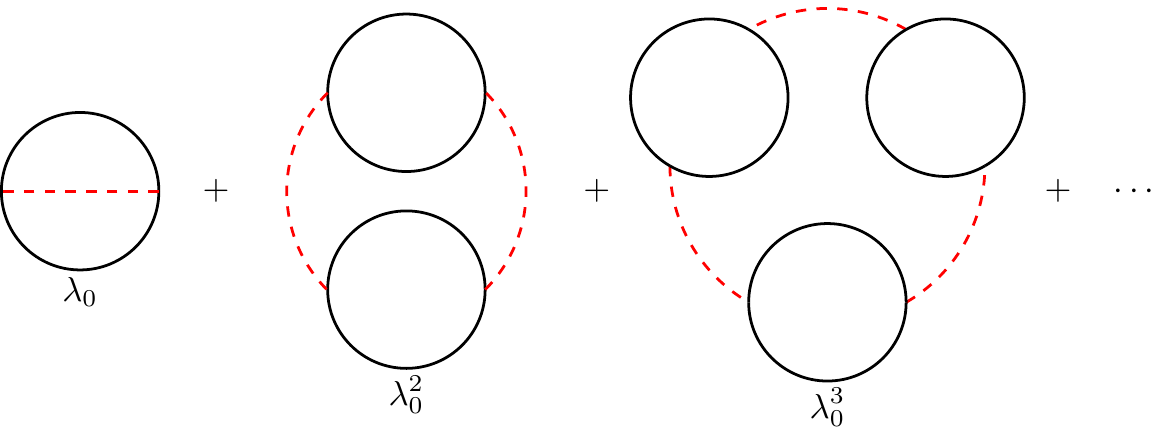}
\caption{Sum of the ring diagrams contributing to $V_{(1)}(\sigma_{(0)}, \chi_{(0)};h)$.}
\label{nlsmrings}
\end{figure}

As explained in \cite{root} (see also \cite{mr-new}), the next-to-leading 
correction to the effective potential in the $1/N$ expansion is given by a sum of ring diagrams. A ring diagram at $m$ loops is constructed with $m$ bubbles of $\eta$ particles successively connected by $m$ propagators of $\tilde\chi$ particles until the diagram closes on itself. Each bubble comes with a factor $1/\Delta = N-2$ (the number of particles contributing to the bubble), which cancels with the factor $\Delta$ coming from the pair of vertices that connect with the $\tilde\chi$ propagators. The sum of such ring diagrams is shown in \figref{nlsmrings}. Following the structure used in \cite{mr-new}, the contribution of these diagrams is the infinite sum
\begin{equation}
-\sum_{m\geq 1} \frac{1}{2 m}  \int\frac{\rd^d k}{(2\pi)^d} \Big(2 \pi \lambda_0 D_{\tilde{\chi}\tilde{\chi}}(k) \Pi(k^2,h^2+\chi)\Big)^m,
\label{ring}
\end{equation}
where
\begin{equation}
\Pi(k^2,M^2) ={1\over 2} \int {\rd^d q \over (2\pi)^d} {1\over (q^2 + M^2) \left( (k+q)^2 + M^2\right)}
\end{equation}
 is the scalar polarization loop arising from the $\eta$ bubbles.
 
The momentum integrals in \eqref{ring} are divergent. However, going back to \eqref{v-leading-order}, after renormalization of the term with $1/\lambda_0$, the renormalization constant at next-to-leading  order $Z_{(1)}^{-1}$ provides additional divergent terms that should cancel the divergences of our momentum integrals. By evaluating at the critical point with $\chi_{(0)}=0$, we find the renormalized free energy at next-to-leading order
\be
\label{renV1}
\delta \CF_{(1)}(h) = -{h^2 \nu^{-\epsilon} \over 4 \pi } \left[{Z_{(1)}^{-1} \over \lambda}+  \frac{2 \pi \nu^\epsilon}{h^2}  \sum_{m \ge 1} {(-1)^{m} \over m}
 \left({2 \pi \lambda_0 \over \sigma_{(0)}^2} \right)^m \CI_m\right]. 
\ee
The integrals $\CI_m$ are defined as
\be
\CI_m= \int \frac{\mathrm{d}^d k}{(2\pi)^d}\left[ \frac{\Pi(k^2, h^2)(k^4+4h^2k_0^2)}{k^2}\right]^m.
\label{eq_V1_Im}
\end{equation}
By first expanding $(k^4+4h^2k_0^2)^m$ with the binomial theorem, the angular integral of the momentum can be computed term by term. $\CI_m$ can then be expressed as a finite sum of rotational invariant integrals. Finally, to compute the integral over $k^2$, it will be crucial to write the scalar polarization function in terms of a hypergeometric function:
\be
\Pi(k^2, h^2)= h^{-2-\epsilon} {\Gamma(1+ \frac{\epsilon}{2}) \over 2 (4 \pi)^{d/2}} 
 \left( {x+4 \over 4} \right)^{-1-\epsilon/2} {}_2 F_1 \left( 1+{\epsilon \over 2}, {1\over 2}; {3\over2}; {x \over x+4} \right). 
\ee
In this equation, 
\be
x={k^2 \over h^2}. 
\ee
The integrals $\CI_m$ then can be expressed as
\begin{equation}
\CI_m = h^{2-\epsilon(m+1)} \frac{S_d}{2(2\pi)^d} \left[ \frac{\Gamma(1+\frac{\epsilon}{2})}{2(4\pi)^{d/2}} \right]^m \sum_{\ell=0}^m \binom{m}{\ell} \frac{(1/2)_\ell}{(d/2)_\ell} 4^\ell \CI_{m,\ell},
\label{Im}
\end{equation}
where $(x)_n$ is the Pochhammer symbol, $S_d=2 \pi^{d/2}/\Gamma(d/2)$ is the volume of a $d$-dimensional sphere, and 
\be
\label{Iml}
\CI_{m, \ell}= 4^{d/2+m-\ell} \int_0^1 \rd z\, {z^{m-\ell-\epsilon/2} \over (1-z)^{2-\ell-(m+1)\epsilon/2}} \left[ {}_2 F_1 \left( 1+{\epsilon \over 2}, {1\over 2}; {3\over2}; z \right) \right]^m, 
\ee
which we have appropriately expressed in terms of the integration variable $z = x/(x+4)$. It is easy to see that, when $\ell\ge 2$, the integrals $\CI_{m, \ell}$ are finite as $\epsilon \rightarrow 0$. Thus, for now we focus on the integrals with $\ell=0$, $1$ and their singular part.

It is convenient to re-express these integrals in yet another form, specially in order to extract the singular part. We use fractional linear transformations of the hypergeometric function to write
\begin{multline}
{}_2 F_1 \left( 1+\frac{\epsilon}{2}, {1\over 2}; {3\over2}; z \right) = -\frac{1}{\epsilon} \frac{\Gamma(1-\frac{\epsilon}{2})^2}{2^\epsilon\,\Gamma(1-\epsilon)} z^{-1/2}\\
\times\left[ 1 - \frac{2^\epsilon\,\Gamma(1-\epsilon)}{\Gamma(1-\frac{\epsilon}{2})^2} (1-z)^{-\epsilon/2} {}_2 F_1 \left(-\frac{\epsilon}{2}, \frac{1}{2}; 1-{\epsilon \over2}; 1-z \right) \right].
\label{2F1-transform}
\end{multline}
Similar manipulations of the polarization loop were done in \cite{abe,ma} to calculate critical exponents in the $1/N$ expansion. We can now write the integral $\CI_{m, \ell}$ as
\begin{align}
\CI_{m, \ell} &= 4^{d/2+m-\ell} \left(  -\frac{1}{\epsilon} \frac{\Gamma(1-\frac{\epsilon}{2})^2}{2^\epsilon\,\Gamma(1-\epsilon)} \right)^m \CB_{m, \ell}, 
\label{Iml-bml}\\
\CB_{m, \ell} &= \int_0^1 \rd z \, { (1-z)^{m/2 -\ell-\epsilon/2}  \over z^{ 2-\ell -(m+1) \epsilon/2}} \left[ 1 - \frac{2^\epsilon\,\Gamma(1-\epsilon)}{\Gamma(1-\frac{\epsilon}{2})^2} z^{-\epsilon/2}
{}_2 F_1 \left(-{\epsilon \over 2}, {1\over 2}; 1-{\epsilon \over2}; z \right)  \right]^m.
\label{bml}
\end{align}
%\be
In writing this expression, we have changed the integration variable from $z$ to $1-z$. The integral \eqref{bml} can be written as an infinite sum which will be useful for our analysis. Let us define the expansion coefficients
\be
\left[{}_2 F_1 \left(-{\epsilon \over 2}, {1\over 2}; 1-{\epsilon \over2}; z \right)  \right]^s=  \sum_{k \ge 0} c_k^{(s)} z^{ k}.
\label{2F1-exp}
\ee
Then, by expanding the binomial in \eqref{bml} and using \eqref{2F1-exp} for each of the hypergeometric functions arising in the binomial sum, we can integrate term by term in $z$, and we find:
\be
\CB_{m,\ell}= \sum_{k \ge 0} \CB_{m,\ell, k},
\label{bml-exp}
\ee
where 
\begin{equation}
\label{bmlk}
\CB_{m,\ell, k} = \sum_{s=0}^m \binom{m}{s} \left(-\frac{2^\epsilon\,\Gamma(1-\epsilon)}{\Gamma(1-\frac{\epsilon}{2})^2}\right)^s \frac{\Gamma \left(-\ell+\frac{m}{2}-\frac{\epsilon }{2}+1\right) \Gamma
\left(k+\ell+(m-s+1) \frac{\epsilon}{2} -1\right)}{\Gamma
\left(k+\frac{m}{2}+(m-s) \frac{\epsilon}{2} \right)} c_k^{(s)}.
\end{equation}
   %
%In our result \eqref{bmlk}--\eqref{bml-exp} we have successfully isolated the full singularity of $\CI_{m,\ell}$ in the first two terms $k=0$, $1$. From the factor $1/\epsilon^m$ in \eqref{Iml-bml}, we note that the singular part of $\CI_{m,\ell}$ corresponds to the Laurent expansion of $\CB_{m,\ell}$ up to order $\epsilon^{m-1}$. Thus, the first two terms $k=0$, $1$ fully contain the expansion of $\CB_{m,\ell}$ up to this order. We prove this statement in appendix \ref{computation_bhat}.
In appendix \ref{computation_bhat}, we prove that the terms $k=0$, $1$ in \eqref{bmlk} contain the Laurent expansion of $\mathcal{B}_{m,\ell}$ up to order $\epsilon^{m-1}$. From the factor $1/\epsilon^m$ in \eqref{Iml-bml}, we note that the expansion of $\mathcal{B}_{m,\ell}$ up to this order precisely corresponds to the singular part of $\CI_{m,\ell}$. Thus, the two terms $\CB_{m,\ell, 0}$ and $\CB_{m,\ell, 1}$ fully contain the singularities of $\mathcal{I}_{m,\ell}$.

\subsection{Divergences and beta function}
\label{beta-singular-part}

The renormalized free energy \eqref{renV1} must be finite. By imposing cancellation 
of divergences we should be able to obtain an explicit expression for 
$Z_{(1)}^{-1}$, and thus, for the next-to-leading 
order of the beta function in the $1/N$ expansion. This result, which we quoted in \eqref{beta-bh-1}, has been known for some time \cite{bh}, based on the $1/N$ calculation of critical exponents \cite{abe, ma}. Our calculation provides a 
direct derivation of the beta function, very similar to the calculation by Palanques-Mestre and Pascual in \cite{pmp}, 
where they studied the beta function of QED in the $1/N_f$ expansion.

Let us write the next-to-leading correction to $Z^{-1}$ in (\ref{largeZ}) as a Laurent expansion with yet undetermined coefficients $B_i^{(n)}$:
\be
\label{z1N}
Z_{(1)}^{-1}=\sum_{n \ge 1} \lambda^n \sum_{i=0}^{n-1} { B_i^{(n)} \over \epsilon^{n-i}}. 
\ee
Firstly we will relate the coefficients $B_i^{(n)}$ directly with the coefficients of the beta function and, secondly, we will find an explicit expression for $B_i^{(n)}$ by imposing cancellation of divergences in \eqref{renV1} and using our result in \eqref{bmlk}.

The beta function for the 't Hooft coupling can be obtained from the renormalization constant $Z^{-1}$ as
\be
\beta(\lambda)= -\epsilon { \lambda \over 1- \lambda {\partial \over \partial \lambda} \log Z^{-1}}. 
\ee
By using  the Laurent expansion \eqref{z1N} and the leading order result \eqref{Z0} for the renormalized coupling, the 
above equality can be written at order $\Delta$ in the $1/N$ expansion as
\begin{equation}
\beta_ {(1)}(\lambda) =- \lambda^2 \sum_{n \ge 0} \lambda^n (n+1) B_n ^{(n+1)}.
\label{b1-Bi-relation}
\end{equation}
In addition, finiteness of the $\beta$ function as $\epsilon \rightarrow 0$ requires that
\be
\label{Bcons}
B_i^{(n+1)}= -{n-1 \over n+1}  B^{(n)}_i, \qquad i=0, \dots, n-1, \quad n \ge 1. 
\ee
Thus, the computation of $\beta_{(1)}(\lambda)$ now reduces to determining the coefficients $B_i^{(n)}$. 

As we prove in appendix \ref{computation_bhat}, the divergent part of the integrals $\CI_{m,\ell}$ comes from a finite number of terms $k$ in the sum of \eqref{bml-exp}. To be more specific, $\CB_{m,\ell,k}$ leads to singularities only for $k=0$, $1$ when $\ell=0$, and for $k=0$ when $\ell=1$. Thus, the divergent part of $\CI_{m}$ will be contained completely in the sum
 \be
 \label{SB}
 \CS_m= \CB_{m,0,0}+ \CB_{m,0,1} + {m \over 2 -\epsilon} \CB_{m,1,0} ,
 \ee
 where the factor in front of the $\ell=1$ term arises from the binomial coefficient and Pochhammer symbols in \eqref{Im}. By combining appropriately the $\Gamma$ factors in \eqref{bmlk}, one finds 
  \be
 \label{com-sun-binomial}
\CS_m={2 \over \epsilon} \Gamma\left( {m \over 2}-{\epsilon \over 2}\right) \sum_{s=0}^m \binom{m}{s} {(-1)^s \over m-s+1} f_m (s\epsilon,\epsilon),
   \ee
where
\begin{equation}
\label{doublef}
\begin{aligned}
f_m (y, x) &= { \Gamma( {m+1 \over 2} x -{y \over 2}  +1)
\over \Gamma( {m (1+x) \over 2}-{y\over 2})} \left( \frac{m-x }{(m+1)x -y-2}+\frac{y
(m-x )}{2 (x -2) (m (x+1)-y )}+\frac{m}{2-x } \right) \re^{y \mathfrak{g}(x)},\\
\mathfrak{g}(x) &= {1\over x} \log \left( \frac{2^x\, \Gamma(1-x)}{\Gamma(1-\frac{x}{2})^2}\right).
\end{aligned}
\end{equation}
By Taylor expanding $f_m (s\epsilon, \epsilon)$ in powers of the first variable and using properties of the binomial coefficient, we can compute the sum in \eqref{com-sun-binomial}. We obtain, 
\begin{equation}
\label{divpart}
\CS_m=  {(-1)^m \over m+1} \left(\frac{2^\epsilon\,\Gamma(1-\epsilon)}{\Gamma(1-\frac{\epsilon}{2})^2}\right)^{m+1}
{\epsilon-1 \over \epsilon-2}-\Gamma\left({m \over 2} \right) {2(-1)^{m}  \over m+1} \left[ \frac{\rd^{m+1} f_m(y,0)}{\rd y^{m+1}}\right]_{y=0} \epsilon^m+\CO\left(\epsilon^{m+1}\right).
\end{equation}
We remark that the expression we have obtained is only valid up to order $\epsilon^m$, while the first term in the r.h.s. is equal to $\CS_m$ up to order $\epsilon^{m-1}$ (so it correctly encodes the singular part of the integrals $\CI_m$). We will however need the second term to extract the finite part of the free energy.
 
We are now ready to extract the singular part of the sum over ring diagrams in \eqref{renV1}. First, let us introduce some notation. Let $A(\epsilon)$ be a Laurent series in $\epsilon$. We will denote by $\div A(\epsilon)$ the singular or principal part of the series. Then, it is easy to show that 
 \be
 \label{jpi}
\div \left[ \frac{2 \pi \nu^\epsilon}{h^2}  \sum_{m \ge 1} {(-1)^{m} \over m}
 \left({2 \pi \lambda_0 \over \sigma_{(0)}^2} \right)^m  \CI_m  \right] = \div \left[ \sum_{r\ge 1} \lambda_0^r \nu^{-r \epsilon} \Pi_r (h; \epsilon) \right], 
\ee
where
\begin{equation}
\Pi_r (h;\epsilon) = H^{-\epsilon(r+1)} { 2^{1-\epsilon}  \over \epsilon^r }{\left[ \Gamma(1+{\epsilon\over 2}) \right]^r 
\over \Gamma\left(1-{\epsilon\over 2} \right)} \sum_{p=0}^{r-1} {1\over r-p} \binom{r-1}{p} 
\left(\frac{\Gamma(1-\frac{\epsilon}{2})^2}{2^\epsilon\,\Gamma(1-\epsilon)}\right)^{r-p}  \CS_{r-p},
\label{pi-r}
\end{equation}
and we have denoted
\begin{equation}
H^2 = \frac{h^2}{4\pi\nu^2}. 
\end{equation}
The expression in \eqref{jpi} is obtained by using \eqref{sigma0} in place of $\lambda_0/\sigma_{(0)}^2$ and 
reexpanding in powers of $\lambda_0$. Replacing $\CS_m$ in \eqref{pi-r} by the first term of \eqref{divpart}, we find
\be
\div\big[ \Pi_r(h; \epsilon) \big] = \div\left[  {1\over (r+1) \epsilon^r} P\left( \epsilon, (r+1) \epsilon \right) \right], 
\ee
where 
\begin{equation}
\label{doubleP}
P(x, y) = -2 H^{-y} \frac{\re^{(y-x) \mathfrak{j}(x)}}{\Gamma(1-\frac{x}{2})} \frac{\Gamma(1-x)}{\Gamma(1-\frac{x}{2})^2} {x-1 \over x-2}, \qquad 
 \mathfrak{j}(x)={1\over x} \log\Gamma\left(1+{x \over 2} \right). 
\end{equation}
 We can now use a similar argument to the one in \cite{pmp}. In the r.h.s. of \eqref{jpi}, we replace 
 $\lambda_0$ by the renormalized coupling at leading order in $1/N$: 
 \be
 \lambda_0 = { \nu^{\epsilon} \lambda \over 1 + {\lambda \over \epsilon}}. 
 \ee
 We then obtain, 
 \be
 \label{final-ser}
 \sum_{r\ge 1} \frac{\lambda^r}{(1 + \frac{\lambda}{\epsilon} )^r}\Pi_r (h; \epsilon)  =  \sum_{m\ge 1} \left[ {(-1)^{m+1} \over m(m+1)} {P_0(\epsilon) \over \epsilon^{m}}+ (m-1)! P_m(\epsilon) +\CO\left(\epsilon\right) \right] \lambda^m ,
  \ee
  where we used that
  \be
  \sum_{s=0}^{m-1} \binom{m-1}{s} (-1)^s  (m+1-s)^{j-1}=
  \begin{cases}
  \dfrac{(-1)^{m+1}}{m(m+1)}, & j=0,\\
  0, &  j=1, \dots, m-1,\\
  (m-1)!, & j=m,
  \end{cases}
  \ee
  and we expanded
\be
\label{expandP}
P(x,y)= \sum_{j=0}^\infty P_j(x) y^j.
\ee
Because the expansion functions $P_j(x)$ are regular at $x=0$, it is obvious from \eqref{final-ser} that only $P_0(\epsilon)$ contributes to the singular part of the sum over ring diagrams. By using the reflection formula for the gamma function, we obtain the explicit expression 
\be
\label{p0ex}
P_0(\epsilon) = P(\epsilon,0) = - \frac{4\sin\left(\frac{\pi \epsilon}{2}\right)}{\pi \epsilon}\frac{\Gamma(1-\epsilon)}{\Gamma(1-\frac{\epsilon}{2})^2} \frac{\epsilon - 1}{\epsilon - 2},
\ee
which has a power series expansion at $\epsilon=0$ of the form:
\begin{equation}
P_0(\epsilon) =  \sum_{i \ge 0} P_{0,i} \epsilon^i.
%= - 1 +{\epsilon \over 2} +{\epsilon^2 \over  4} +\epsilon^3 {1-2 \zeta(3) \over 8 } +\epsilon^4{60 \zeta(3) -\pi^4 + 30 \over 480}+ \cdots. 
\end{equation}
Requiring cancellation of divergences in \eqref{renV1} determines the expansion of \eqref{z1N}, and we find the values
\be
B_i^{(m)}= {(-1)^{m-1} \over m(m-1)} P_{0,i-1}, \qquad m \ge 2, \quad i \ge 1, 
\ee
as well as
\be
\qquad B_0^{(m)} = 0, \qquad m \ge 1, 
\ee
which satisfy the constraint \eqref{Bcons}. The beta function at next-to-leading order in the $1/N$ expansion can be now computed by going back to \eqref{b1-Bi-relation} and using our result for the coefficients $B_i^{(n)}$:
\be
\label{betaex}
\beta_{(1)}(\lambda) =-\lambda^2 \sum_{m \ge 1 } {(-1)^{m} \over m} P_{0,m-1} \lambda^m = \lambda^2 \int_0^\lambda P_0(-x) \rd x.
%&=-\lambda^3 \left\{ 1+\frac{\lambda }{4}-\frac{\lambda ^2}{12}+\frac{ \lambda ^3}{32}
   %(1-2 \zeta (3))+\frac{\lambda ^4 \left(-60
   %\zeta (3)-30+\pi ^4\right)}{2400}+\CO\left(\lambda ^5\right)\right\}.
\ee
This result of course coincides with \eqref{beta-bh-1}.

\subsection{Finite part of the free energy and comparison with the Bethe ansatz}

Once the singularities have been canceled, we can focus on the finite part of the renormalized free energy. The finite part arises from 
the integrals $\CI_{m, \ell}$ with $\ell \ge 2$ and from the finite part of the integrals $\CI_{m, \ell}$ with $\ell=0,1$. 
This last contribution comes from three types of terms. Two of them are already written down in the previous section, since they have their origin in $\CS_m$: the last term in the r.h.s. of 
\eqref{divpart}, and the finite part in the r.h.s. of \eqref{final-ser}. In addition, we have to take into account the contribution from the terms 
\be
\widehat \CB_{m,0}=\sum_{k \ge 2} \CB_{m, 0, k}, \qquad  \widehat \CB_{m,1}= \sum_{k \ge 1} \CB_{m,1, k },
\label{bhat}
\ee
which are not included in \eqref{SB}. It will be convenient to rewrite the series \eqref{bhat} in the following integral representation, which we derive in appendix \ref{computation_bhat}:
\be
\label{habmell}
\widehat \CB_{m, \ell} = \epsilon^m (-1)^m \int_0^1 \mathrm{d}z \frac{(1-z)^{m/2-\ell}}{z^{2-\ell}} \left[ \frac{\mathrm{d}^m g_{\ell}(y;z)}{\mathrm{d}y^m}\right]_{y=0} + \CO\left(\epsilon^{m+1}\right), 
\end{equation}
where
\be
\label{fell}
g_{\ell}(y;z) = \left(\frac{\sqrt{z}}{2}\right)^{-y} \left[ \left(\frac{1+\sqrt{1-z}}{2}\right)^y - 1 + \delta_{\ell 0}\frac{zy}{4} \right], \qquad \ell=0,1. 
\ee
The derivation of this result relies on the same tricks that we used to obtain \eqref{divpart} (similar manipulations can also be found in \cite{pm}).

We can now write the next-to-leading correction to the renormalized free energy as 
\be
\delta\CF_{(1)}(h)= - {h^2 \over 2} \left\{ W\left(\lambda; {h\over \mu} \right)+ X\left(\lambda; {h\over \mu} \right) + Y\left(\lambda; {h\over \mu} \right)+ Z\left(\lambda; {h\over \mu} \right) \right\}, 
\ee
where
\be
\label{def_WXYZ}
\ba
W\left(\lambda; {h\over \mu} \right)&= {1\over 2 \pi} \sum_{m\ge 1} \left[ \frac{(-1)^{m+1}}{m(m+1)}P_{0,m} + (m-1)! P_m(0)\right] \lambda^m ,\\
X\left(\lambda; {h\over \mu} \right)&= {1\over  \pi} \sum_{m \ge 1} {1\over m}\lim_{\epsilon\rightarrow 0} \frac{1}{\epsilon^m} 
\left(\widehat \CB_{m,0}+ {m \over 2} \widehat \CB_{m,1}\right) \left( {\lambda\over 1+ \lambda\log(h/\mu)} \right)^m, \\
Y\left(\lambda; {h\over \mu} \right) &= -\frac{2}{\pi} \sum_{m\ge 1} \frac{(-1)^m}{m(m+1)}\Gamma\left( \frac{m}{2} \right) \left[ \frac{\rd^{m+1} f_m(y,0)}{\rd y^{m+1}}\right]_{y=0} \left( {\lambda\over 1+ \lambda\log\left(h/\mu \right)} \right)^m,\\
Z\left(\lambda; {h\over \mu} \right)&= \frac{1}{4\pi}\sum_{m\geq 1} \frac{(-1)^m}{m} \sum_{\ell=2}^m \binom{m}{\ell}\frac{(1/2)_\ell}{\ell!}4^{\ell-m}\CI_{m, \ell} \left( {\lambda\over 1+ \lambda\log\left(h/\mu \right)} \right)^m. 
\ea
\ee
Many of the ingredients appearing in these formulae have been already spelled out in detail. The coefficients $P_{0,m}$ can be read from \eqref{p0ex}. 
The coefficients $P_m(0)$ follow from \eqref{doubleP} and \eqref{expandP}:
\be
P_m(0)={(-1)^{m+1} \over m!} \log^m\left( {h \over \mu} \right). 
\label{pm(0)}
\ee
The function $f_m (y,x)$ is given in \eqref{doublef}. It remains to compute the integrals $\widehat\CB_{m,0}$, $\widehat\CB_{m,1}$ and $\CI_{m,\ell}$, $\ell=2, \dots, m$, for arbitrary $m, \ell$. This can be done analytically, and the results are presented in appendix \ref{finite-ints}. This allows us to compute $\delta \CF_{(1)}(h)$ at any given order. The very first terms read
\begin{multline}
\delta \CF_{(1)}(h) = -\frac{h^2}{4 \pi}\bigg[
\left( -\frac{1}{4} + \log \left(h/\mu\right)\right) \lambda
+ \left(- \frac{7}{24} + \frac{21 \zeta (3)}{32} + \frac{\log \left(h/\mu\right)}{2} - \frac{\log ^2\left(h/\mu\right)}{2}\right) \lambda^2\\
+ \left( - \frac{23}{96} - \frac{107 \zeta (3)}{96} + \frac{(8 - 21\zeta(3))\log \left(h/\mu\right)}{16} - \frac{\log ^2\left(h/\mu\right)}{2} + \frac{\log^3\left(h/\mu\right)}{3}\right)\lambda^3\\
+ \biggl( - \frac{121}{320} + \frac{3659 \zeta (3)}{2560} + \frac{3\zeta(4)}{320} + \frac{4185 \zeta (5)}{2048} + \frac{(24 + 105 \zeta (3)) \log \left(h/\mu\right)}{32}\\
+ \frac{(-24 + 63\zeta(3)) \log ^2\left(h/\mu\right)}{32} + \frac{\log^3\left(h/\mu\right)}{2} - \frac{\log ^4\left(h/\mu\right)}{4} \bigg)\lambda^4 + \CO\left(\lambda^5\right) \bigg].
\end{multline}
In order to compare with the Bethe ansatz solution, we have to rexpress this result in terms of the coupling constant $\alpha$, defined in \eqref{a-tilde}. The first step is 
to set $\lambda$ to the running coupling constant at the scale $\mu=h$, which defines $\widebar\lambda = \lambda\left(\mu=h \right)$. $\widebar\lambda$ is related 
to the dynamically generated scale $\Lambda$ and $h$ by 
\begin{equation}
\log\left(\frac{h}{\Lambda}\right) = \frac{1}{
\bar{\lambda}}+\xi \log(
\bar{\lambda})+\int_0^{\bar{\lambda}} \left[\frac{1}{\beta(u)}+\frac{1}{u^2}-\frac{\xi}{u}\right]\rd u
\label{bart_from_h}, 
\end{equation}
where $\xi$ is defined in (\ref{xi-def}). At this scale, $\delta \CF_{(1)}(h)$ simplifies greatly to
\be
\ba
\delta \CF_{(1)}(h)&=-\frac{h^2}{4 \pi}\sum_{m\ge 1} v_m \widebar \lambda^m
= -\frac{h^2}{4 \pi}\left[-\frac{\widebar \lambda}{4}+ \left(-\frac{7}{24} + \frac{21 \zeta (3)}{32}\right) \widebar\lambda^2 + \cdots\right]. 
\ea
\ee
This defines the coefficients $v_m$, $m \ge 1$. We can now use the Legendre transform (\ref{legendre}) to obtain the normalized 
energy density (\ref{ealpha}). In order to do that, it is useful to introduce yet another coupling 
\be
\frac{1}{\tilde{\alpha}}+\xi \log\tilde{\alpha} = \log\left(\frac{h}{\Lambda}\right),
\label{alphat_from_h}
\end{equation}
which was first considered in \cite{bbbkp} and is related to $\bar \lambda$ by
\begin{equation}
\bar{\lambda}= 
\tilde{\alpha}
-\Delta
\left(\tilde{\alpha}^2\int_0^{\tilde{\alpha}} \left[\frac{1}{u}+\frac{\beta_{(1)}(u)}{u^4}\right]\rd u\right) +\mathcal{O}\left(\Delta^2\right).
\end{equation}
We can use $\tilde{\alpha}$ to write the free energy as
\begin{equation}
\delta \CF(h) =  h^2 \left(\frac{1}{\Delta}\Sigma_0(\tilde{\alpha})+\Sigma_1(\tilde{\alpha}) + \CO(\Delta) \right),
\label{refdef}
\end{equation}
where
\begin{equation}
\ba 
\Sigma_0(\tilde{\alpha}) &= - \frac{1}{4\pi}\left( \frac{1}{\tilde{\alpha}}-\frac{1}{2}\right),\\
\Sigma_1(\tilde{\alpha}) &=  -\frac{1}{4\pi}\sum_{m\geq 1} v_m \tilde{\alpha}^m
- \frac{1}{4\pi}\int_0^{\tilde{\alpha}} \left[\frac{1}{u}+\frac{\beta_{(1)}(u)}{u^4}\right]\rd u.
\ea
\end{equation}
The Legendre transform gives, 
\be
\ba
 {\rho\over h} &=\frac{1}{2\pi\Delta \tilde{\alpha}}+\frac{\tilde \alpha}{4\pi}-2\Sigma_1(\tilde{\alpha})+\tilde\alpha^2\Sigma'_1(\tilde{\alpha})+\CO(\Delta),\\
 {e\over h^2} &= \frac{1}{\Delta }\left(\frac{1}{4 \pi  \tilde{\alpha}}+\frac{1}{8 \pi }\right)+\frac{\tilde{\alpha} }{4 \pi }-\Sigma _1(\tilde{\alpha})+\tilde{\alpha}^2 \Sigma _1'(\tilde{\alpha})+\CO\left(\Delta \right). 
 \ea
 \ee
 The final step is to relate $\tilde \alpha$ to the coupling $\alpha$ defined in (\ref{a-tilde}):
 \be
 \tilde{\alpha} = \alpha+ \frac{2\pi \Delta \alpha ^3}{ \alpha +1} \left(\frac{\alpha}{4\pi}-2  \Sigma _1(\alpha )+  \alpha ^2 \Sigma _1'(\alpha ) \right)+\CO\left(\Delta^2\right),
\end{equation}
which leads to a remarkably simple expression for the normalized energy density:
\begin{equation}
\frac{e}{\rho^2\pi\Delta} =\alpha+\frac{\alpha ^2}{2} -   \Delta  \left(\alpha ^2\sum_{m\geq 1} v_m \alpha^m
+ \alpha^2\int_0^{\alpha} \left[\frac{1}{u}+\frac{\beta_{(1)}(u)}{u^4}\right]\rd u\right)+\CO\left(\Delta ^2\right)\,.
\label{efrompot}
\end{equation}
Expanding $\beta_{(1)}(u)$ with \eqref{betaex}, we can read the following result for the series $\CE_{(1)}(\alpha)$, defined in \eqref{ela}, in terms of the perturbative coefficients $v_m$:
\be
\label{e1per}
\CE_{(1)}(\alpha)= -\alpha^2  \sum_{m \ge 1} \left( v_m  -{(-1)^{m+1} \over m(m+1)} P_{0,m}\right) \alpha^{m}. 
\ee
It follows from this expression that 
\be
\label{exyz}
 \CE_{(1)}(\alpha) = -2\pi\alpha^2\big[ X(\alpha;1)+ Y(\alpha;1)+ Z(\alpha;1) \big],
 \ee
 where the functions in the r.h.s. were defined in \eqref{def_WXYZ}. 
 
%One can now compare the expression \eqref{e1per} with the result of the Bethe ansatz \eqref{e1-ba}. We find perfect agreement up to order $\alpha^{44}$. Up to this order, we find that every coefficient is the sum of a rational number plus a linear combinations of odd Riemann zeta functions\footnote{From the Bethe ansatz result, we also know that higher corrections in $\Delta$ can be written as polynomials in multiple variables of odd Riemann zeta functions.}. In appendix \ref{finite-ints} we prove to all orders that $Z(\alpha;1)$ can be written as linear combination of $\zeta(2k+1)$ (in fact, with no rational term). On the other hand, $X(\alpha;1)$ and $Y(\alpha;1)$ do not satisfy this transcendentality property when alone, but their combination indeed does, up to order $\alpha^{44}$ (in this case, a rational term has to be included with the linear combination of odd zetas), although we do not have a proof of this statement to all orders.

One can now compare the expression \eqref{e1per} with the result of the Bethe ansatz \eqref{e1-ba}. We find perfect agreement up to order $\alpha^{44}$.

As a side remark, up to order $\alpha^{44}$, we notice that every coefficient of $\CE_{(1)}(\alpha)$ is the sum of a rational number plus a linear combinations of odd Riemann zeta functions\footnote{From the Bethe ansatz result, we also know that higher corrections in $\Delta$ can be written as polynomials in multiple variables of odd Riemann zeta functions.}. In appendix \ref{finite-ints} we prove to all orders that $Z(\alpha;1)$ can be written as linear combination of $\zeta(2k+1)$ (in fact, with no rational term). On the other hand, $X(\alpha;1)$ and $Y(\alpha;1)$ do not satisfy this transcendentality property when alone, but the combination $X(\alpha;1) + Y(\alpha;1)$ indeed does, up to order $\alpha^{44}$ (in this case, a rational term has to be included with the linear combination of odd zetas), although we do not have a proof of this statement to all orders.
%
%\be
%\CE

\section{Large \texorpdfstring{$N$}{N} renormalons and their trans-series}
\label{ren-sec}

In \cite{volin,mr-ren}, numerical evidence was given for the factorial growth of the 
perturbative series for \eqref{epsrho}, at fixed $N$. This was interpreted as a signature of renormalons \cite{beneke,parisi1,parisi2}. In \cite{mr-ren}, the contribution of UV and IR renormalons was disentangled, and detailed evidence was given that the large order behavior 
of the perturbative series is in agreement with the predictions of renormalon physics. In particular, it was 
shown that the next-to-leading behavior of the asymptotics involves the first two 
coefficients $\beta_0$ and $\beta_1$ of the beta function. The evidence for these effects was based on a numerical study of the perturbative series and it focused on the leading singularities in the Borel plane. 

At large $N$, the ring diagrams studied in the previous section should give the leading 
renormalon behavior. One advantage of having explicit results for these diagrams is that 
we can obtain from them analytic results on the large order behavior of the perturbative series. Equivalently, we can find explicit results for the exponentially suppressed trans-series associated to each Borel singularity. These can be obtained without even calculating the loop integrals. As shown in \cite{mr-new,mr-hub, mr-roads}, it is enough to write the generating functions 
\eqref{def_WXYZ} in integral form and study their imaginary 
parts (or, equivalently, their imaginary discontinuities). 

We will now present the integral forms for the series $X(\lambda;1)$, $Y(\lambda;1)$ and $Z(\lambda;1)$ appearing in (\ref{exyz}). We note 
that the coefficients of the series $W(\lambda;1)$ do not grow factorially. This is easily observed from \eqref{pm(0)} and the fact that $P_{0,m}$ are the Taylor coefficients of an analytic function at $\epsilon=0$.

Let us start with $X(\lambda;1)$. The integral form in this case is easily obtained by using 
the explicit expression \eqref{fell} and Laplace transforms. We find
\be
\label{Xlambda}
X(\lambda;1)= \frac{1}{\pi}\int_0^1 {\rd z\over z^{2}} \CX_0\big(\lambda\sqrt{1-z},z\big) + \frac{1}{2\pi}\int_0^1 {\rd z \over z (1-z) } \CX_1 (\lambda\sqrt{1-z},z), 
\end{equation}
where
\begin{equation}
\label{x01}
\ba
\CX_0(y,z) &= \log \left[1-y \log \left(\frac{\sqrt{z}}{2}\right)\right] - \log \left[1 - y \log \left(\frac{\sqrt{z}}{1+\sqrt{1-z}}\right)\right] -\frac{y z}{4}\frac{1}{1 - y \log\left(\frac{\sqrt{z}}{2}\right)}, \\
\CX_1(y,z)&= \frac{1}{1- y \log\left(\frac{\sqrt{z}}{1+\sqrt{1-z}}\right)}-\frac{1}{1-y\log\left(\frac{\sqrt{z}}{2}\right)}. 
\ea
\end{equation}

In the case of $Y(\lambda;1)$ we use the explicit expression \eqref{doublef} and write the Euler beta functions that appear in the resulting expression as integrals over $z$. In this way we find 
\be
Y(\lambda;1)=  \frac{1}{4\pi}\int_0^1 \frac{\rd z}{z(1-z)}\sum_{m\geq 1} \frac{(-\lambda\sqrt{1-z})^m}{m+1}
\frac{\rd^{m+1}}{\rd y^{m+1}}
\left[
2^{y} y 
z^{-y/2}\left(\frac{2y}{y+2}
+
z\right)\right]_{y=0}.
\ee
We can use again Laplace transforms to sum up this series and we eventually find
\be
\label{Ylambda}
Y(\lambda;1) = \frac{1}{4\pi}\int_0^1 \frac{\rd z}{z (1-z)} \CY (\lambda\sqrt{1-z}, z), 
\ee
where 
\be
\CY(y,z)=- z + 2z(1-z) + \frac{z\,\re^{-2/y}}{y}\, \text{E}_1\left[- \frac{2}{y} \left( 1 - y\log\left(\frac{\sqrt{z}}{2}\right)\right)\right] + \frac{z+2}{1 - y\log \left(\frac{\sqrt{z}}{2}\right)}
\label{y01}
\ee
and $\text{E}_1(z)$ is the exponential integral. 

Finally, after using the identity
\be
\sum_{m\ge \ell} \frac{1}{m} \binom{m}{\ell}x^m = \frac{x^\ell}{\ell(1-x)^\ell},
\end{equation}
the last series can be written as 
\be
Z(\lambda;1)= \frac{1}{\pi}\int_0^1 {\rd z\over  (1-z)^{2}}\left[ \frac{\CZ(z,\lambda)-1}{2} - 2 \log\left( \frac{1+\sqrt{\CZ(z,\lambda)}}{2} \right) \right],
\label{Zlambda}
\ee
where
\begin{equation}
\CZ(z,\lambda) =  \frac{1+F(z)\lambda}{1+zF(z)\lambda}, \qquad F(z)= {}_2F_1\left(1, \frac{1}{2}; \frac{3}{2}; z \right) = \frac{\tanh^{-1}(\sqrt{z})}{\sqrt{z}}.
\end{equation}

The advantage of the representations \eqref{Xlambda}, \eqref{Ylambda} and \eqref{Zlambda} is that they lead to explicit, exponentially small imaginary terms. These are precisely the trans-series 
associated to the renormalon singularities. 

Let us first consider the function $X(\lambda;1)$.  It has discontinuities when $\lambda<0$, due to the poles and logarithmic branch cut in the integrands of \eqref{Xlambda}. The singularities of the integrand are located at $z_1$ and $z_2$, which are defined by 
\be
\label{def_z12}
\frac{1}{\lambda} =  \sqrt{1-z_1} \log \left(\frac{\sqrt{z_1}}{2}\right), \qquad \frac{1}{\lambda}= \sqrt{1-z_2} \log \left( \frac{\sqrt{z_2}}{1+\sqrt{1-z_2}}\right).
\end{equation}
We find
\begin{multline}
\disc X(\lambda;1) = 2 \pi \ri \Bigg[  \frac{1}{\pi} \int_{z_2}^{z_1} \frac{\rd z}{z^2} - \frac{1}{\pi} \Res\left(\frac{1}{z^2}\CX_0\left(\lambda\sqrt{1-z},z\right),z=z_1\right)\\
 -\sum_{i=1,2} \frac{1}{2\pi} \Res\left(\frac{1}{z(1-z)} \CX_1\left(\lambda\sqrt{1-z},z\right), z=z_i\right) \Bigg]. 
 \label{discX_form}
\end{multline}
This discontinuity can be computed term by term as a trans-series in $\lambda$, i.e. 
as a power series in both $\re^{2/\lambda}$ and $\lambda$. See appendix \ref{app-disc} for details of this computation. We find 
\begin{multline}
\disc X(\lambda;1) = -\ri \bigg[ \left(\frac{4}{\lambda }-\frac{1}{2}\right) \re^{2/\lambda }+\left(\frac{64}{\lambda ^2}+\frac{48}{\lambda }\right) \re^{4/\lambda } +\left(\frac{864}{\lambda ^3}+\frac{1188}{\lambda ^2}+\frac{306}{\lambda }\right) \re^{6/\lambda }\\
+\left(\frac{32768}{3 \lambda ^4}+\frac{63488}{3 \lambda ^3}+\frac{11264}{\lambda ^2}+\frac{1536}{\lambda }\right) \re^{8/\lambda }
+\left(\frac{400000}{3 \lambda ^5}+\frac{330000}{\lambda ^4}+\frac{802000}{3 \lambda ^3}+\frac{80600}{\lambda ^2}+\frac{7100}{\lambda }\right) \re^{10/\lambda }\\
\!+\left(\frac{7962624}{5 \lambda ^6}+\frac{23887872}{5 \lambda ^5}+\frac{5197824}{\lambda ^4}+\frac{2489472}{\lambda ^3}+\frac{501696}{\lambda ^2}+\frac{31632}{\lambda }\right) \re^{12/\lambda } + \CO\left(\re^{14/\lambda }\right) \bigg].
\label{disc_X}
\end{multline}

Let us now consider the function $Y(\lambda;1)$. By investigating the function \eqref{y01}, we see that $Y(\lambda;1)$ has discontinuities both for positive and negative $\lambda$. When $\lambda<0$, there is a pole at $z=z_1$, where $z_1$ was defined in \eqref{def_z12}, and a discontinuity due to the 
exponential integral. For $\lambda>0$, we have a discontinuity due again to the exponential integral, and one finds
\begin{equation}
\begin{aligned}
&\disc Y(\lambda>0;1) = \ri\, \frac{\re^{-2/\lambda}}{2}, \\
&\begin{multlined}
\disc Y(\lambda<0;1) =
\ri \bigg[\left(\frac{4}{\lambda ^2}+\frac{10}{\lambda }\right) \re^{2/\lambda } + \left(\frac{128}{3 \lambda ^3}+\frac{128}{\lambda ^2}+\frac{64}{\lambda }\right) \re^{4/\lambda }\\
+ \biggl(\frac{432}{\lambda ^4}+\frac{1584}{\lambda ^3}+\frac{1476}{\lambda ^2}+\frac{330}{\lambda }\biggr) \re^{6/\lambda }
\!+ \biggl(\frac{65536}{15 \lambda ^5}+\frac{57344}{3 \lambda ^4}+\frac{25600}{\lambda ^3}+\frac{12032}{\lambda ^2}+\frac{1568}{\lambda }\biggr) \re^{8/\lambda }\\
+ \left(\frac{400000}{9 \lambda ^6}+\frac{680000}{3 \lambda ^5}+\frac{1180000}{3 \lambda ^4}+\frac{284000}{\lambda ^3}+\frac{82200}{\lambda ^2}+\frac{7140}{\lambda }\right) \re^{10/\lambda }
+ \biggl(\frac{15925248}{35 \lambda ^7}\\ +\frac{2654208}{\lambda ^6}+\frac{5640192}{\lambda ^5}+\frac{5501952}{\lambda ^4}+\frac{2536704}{\lambda ^3}+\frac{504576}{\lambda ^2}+\frac{31680}{\lambda }\biggr) \re^{12/\lambda } + \CO\left(\re^{14/\lambda }\right) \bigg].
\end{multlined}
\end{aligned}
\label{disc_Y}
\end{equation}

Finally, we consider the discontinuity of $Z(\lambda;1)$, which arises from two sources. The first one is a pole of the integrand, which appears for $\lambda<0$. This occurs at a $z_3$ satisfying 
\be
1+ \lambda z_3 F(z_3)=0. 
 \label{def_z3}
\end{equation}
There is another source of discontinuity due to the square root inside the logarithm, which occurs when $\CZ(z,\lambda)<0$. The discontinuity of this source is given by
\begin{equation}
\frac{4}{\pi}\int_{z_4}^{z_3} \frac{\rd z}{(1-z)^2} \tanh^{-1}\left(\sqrt{\frac{1+ F(z)\lambda}{1+zF(z)\lambda}}\right),
\label{disc}
\end{equation}
where $(z_4,z_3)$ is the subinterval of $(0,1)$ where $\CZ(z,\lambda)$ is negative. The value $z_3$ is the pole previously discussed in \eqref{def_z3}, 
while $z_4$ satisfies
\begin{equation}
1+ \lambda F(z_4) = 0.
\label{def_z4}
\end{equation}
The trans-series obtained in this way is 
\begin{multline}
\disc Z(\lambda;1) =  -\ri \bigg[\left(\frac{6}{\lambda ^2}+\frac{6}{\lambda }\right) \re^{2/\lambda } + \left(\frac{128}{3 \lambda ^3}+\frac{64}{\lambda ^2}+\frac{16}{\lambda }\right) \re^{4/\lambda } + \left(\frac{486}{\lambda ^4}+\frac{756}{\lambda ^3}+\frac{288}{\lambda ^2}+\frac{24}{\lambda }\right) \re^{6/\lambda }\\
+ \biggl(\frac{65536}{15 \lambda ^5}+\frac{8192}{\lambda ^4}+\frac{13312}{3 \lambda ^3}+\frac{768}{\lambda ^2}+\frac{32}{\lambda }\biggr) \re^{8/\lambda }
+ \biggl(\frac{425000}{9 \lambda ^6}+\frac{290000}{3 \lambda ^5}+\frac{193250}{3 \lambda ^4}+\frac{50300}{3 \lambda ^3}\\+\frac{1600}{\lambda ^2}+\frac{40}{\lambda }\biggr) \re^{10/\lambda }
+\biggl(\frac{15925248}{35 \lambda ^7}+\frac{5308416}{5 \lambda ^6}+\frac{4313088}{5 \lambda ^5}+\frac{304128}{\lambda ^4}+\frac{47232}{\lambda ^3}\\+\frac{2880}{\lambda ^2}+\frac{48}{\lambda }\biggr) \re^{12/\lambda } + \CO\left(\re^{14/\lambda }\right) \bigg].
\label{disc_Z}
\end{multline}

\begin{figure}
\centering
\includegraphics{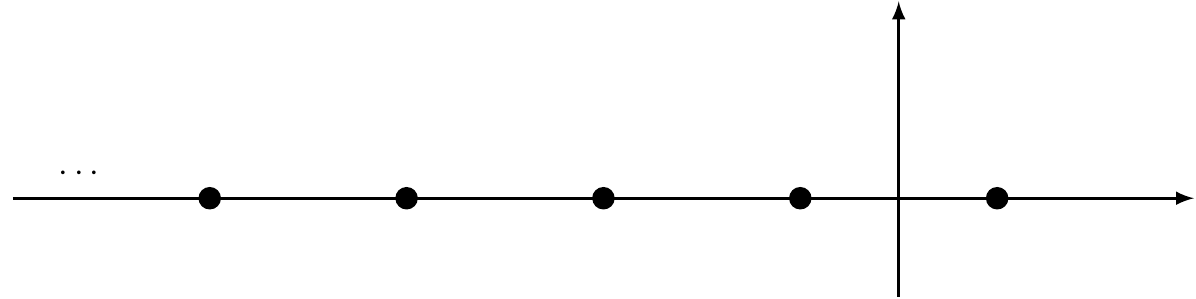}
\caption{Location of the renormalon singularities in the Borel plane. There is a single singularity in the positive real axis for $\zeta=2$, corresponding to the IR renormalon, and and infinite series of UV renormalon singularities at $\zeta= -4k-2$ for $k=0,1,\dots\,$.}
\label{borelplot}
\end{figure}

We can now put all these results together and calculate the trans-series associated to $\CE_{(1)}(\alpha)$. It is given by $1/(2 \ri)$ times 
the discontinuity, and reads
\begin{multline}
\Im \CE_{(1)}(\alpha) = {\pi \over 2}  \biggl[ -\alpha^2 \re^{-2/\alpha} +\left(4 - \alpha ^2\right) \re^{2/\alpha} +\left(\frac{108}{\alpha ^2}+\frac{72}{\alpha }\right)\re^{6/\alpha}\\
+\left(\frac{50000}{9 \alpha ^4}+\frac{20000}{3 \alpha ^3}+\frac{6500}{3 \alpha ^2}+\frac{200}{\alpha }\right) \re^{10/\alpha} + \CO\left(\re^{14/\alpha}\right)  \biggr]. 
\label{disc_E}
\end{multline}
Let us analyze this result. The first term in the r.h.s. of \eqref{disc_E} corresponds to a Borel singularity on the positive real axis at $\zeta=2$. It is an IR renormalon, which has been identified in \cite{volin,mr-ren} at finite $N$ and more recently in \cite{abbh1,abbh2} in the $O(4)$ model. At this order in the $1/N$ expansion there are no additional IR renormalons. The next terms correspond to singularities in the Borel plane on the negative real axis, and they are UV renormalons. We conjecture that they are located at 
\be
\zeta= -4k-2, \qquad k \in \IZ_{\ge 0}. 
\ee
See \figref{borelplot} for a representation of the renormalon singularities in the Borel plane. 
 
From the discontinuity in \eqref{disc_E} we can derive the large order behavior of the coefficients of the series 
\be
\CE_{(1)}(\alpha)=  \sum_{m \ge 3} e_m^{(1)} \alpha^m.
\ee
The derivation follows from the general theory of resurgence (see e.g. \cite{mmbook,abs}), which relates the 
trans-series of a function to the singularities of its Borel transform. In turn, one can extract the large 
order behavior of the perturbative coefficients directly from the Borel singularities. We find
\begin{multline}
\label{em-lo}
e^{(1)}_m = 
-2 \Gamma(m)(-2)^{-m} \left[ 1 - \frac{1}{(m-1)(m-2)} \right]+2\Gamma(m-2)2^{-m}\\
-\frac{3}{2}\Gamma(m+2) (-6)^{-m} \left[ 1 - \frac{4}{m+1} \right]+ \CO \left(m! (-10)^{-m}\right).
\end{multline}
The first term in the r.h.s. gives the leading order asymptotics, which is due to the first UV renormalon singularity at $\zeta = -2$. The second term, with fixed sign, is due to the IR renormalon singularity at $\zeta = 2$; while the third term, which is exponentially subleading with respect to the first two, is due to the next UV renormalon at $\zeta = -6$. 

We can now compare the large order behavior (\ref{em-lo}) with the results of \cite{mr-ren} for finite $N$. In general, one has 
to exercise care in this comparison, since some of the contributions to the asymptotic behavior detected at finite $N$ might be suppressed at large $N$. In \cite{mr-ren} 
it was found that the series (\ref{ealpha}) at finite $N$ has the following asymptotics:
\be
\label{an-asym}
a_n \sim C_{\text{IR}} 2^{-n} \Gamma(n+2 \Delta-1)+C_{\text{UV}} (-2)^{-n} \Gamma(n-2 \Delta+1), \qquad n \gg 1, 
\ee
where the first and second term in the r.h.s. correspond respectively to the IR and UV singularities at $\zeta = \pm 2$, and $C_{\text{IR}}$, $C_{\text{UV}}$ are in principle functions of $N$. After taking into account the difference in the labelling, $m=n+1$, we find that
(\ref{an-asym}) leads to the behavior (\ref{em-lo}) at large $N$, since $\Delta \rightarrow0$ in the arguments of the Gamma functions. This also implies that $C_{\text{IR}}$, $C_{\text{UV}}$ are of order $\Delta$. 

In \cite{mr-ren} it was argued that the IR renormalon 
at $\zeta=2$ is associated to the condensate 
of the operator $\CO= \partial_\mu \boldsymbol{S}\cdot \partial^\mu \boldsymbol{S}$, and that the large order behavior (\ref{an-asym}) is compatible 
with the expectations of renormalon physics. It follows that our large $N$ result for this IR renormalon might be also explained by the contribution 
of this condensate.

The calculation above determines the functional form of the trans-series associated to the different singularities. An additional,
 interesting question is the ``semiclassical decoding" of the normalized density (\ref{ela}) at large $N$, i.e. its expression as a Borel--Ecalle 
resummation of these trans-series and the perturbative series. At finite $N=4$, this has been 
recently done in \cite{abbh1, abbh2} for the full energy density, by using trans-series at finite $N$. Results along this direction in the 
large $N$ expansion 
will be reported in \cite{dmss}.  
%However, it is not clear to us what is the non-perturbative definition of the functions $\CE_{(\ell)}(\alpha)$.
%since as we mentioned in the Introduction, the Bethe ansatz for the non-linear sigma model is not amenable to a $1/N$ expansion. 

\section{The supersymmetric non-linear sigma model}

\label{susy-sec}
We can extend all of the above results to the supersymmetric version of the non-linear sigma model considered in \cite{witten}. 
This model consists of the vector field $\bS(x) $ of the purely bosonic version, satisfying as well the constraint \eqref{cons}, and an $N$-uple 
of two-component Majorana spinors $\bPsi=(\Psi^1, \dots, \Psi^N)$, satisfying the constraint
\be
\label{2cons}
\bS \cdot \bPsi=0. 
\ee
The Lagrangian density is
\be
\label{susyL}
\CL^\text{susy}= \frac{1}{2g_0^2} \bigg\{\partial_\mu \bS\cdot  \partial^\mu \bS + \widebar{\bPsi}\cdot \ri \slashed{\partial}\bPsi + \frac{1}{4}\left( \widebar{\bPsi}\cdot \bPsi\right)^2\bigg\},
\ee
where we follow the conventions of \cite{witten} for the gamma matrices. This model is asymptotically free and its beta function is of the form \eqref{betaf}, with 
\be
\label{susy-beta}
\beta_0= {1\over 4 \pi \Delta}, \qquad \beta_1=0, 
\ee
and $\Delta$ is again given by \eqref{Delta2}. The model can also be studied in the large $N$ expansion, where one finds 
a non-perturbative mass gap and dynamical breaking of the discrete chiral symmetry $\bPsi \rightarrow \gamma^5 \bPsi$ \cite{orlando}. 
The beta function in the $1/N$ expansion has the structure \eqref{betaN}, where $\beta_{(0)}(\lambda)$ is given again by the expression in \eqref{beta-bh-0}, 
while $\beta_{(1)}(\lambda)=0$ due to cancellations between bosons and fermions \cite{gracey-super}. We will rederive this result in section \ref{sec_susysing}. 

As in the non-linear sigma model, we couple the present model to an external potential by using again the conserved charge
$Q^{12}$ associated to the global $O(N)$ symmetry. The dependence of the ground state energy 
on the external potential can be obtained from the Bethe ansatz and the exact $S$-matrix conjectured in \cite{sh-witten}. The resulting 
integral equation was written down explicitly in \cite{eh-ssm}, where it was used to obtain the exact mass gap of the model. In 
\cite{mr-ren} the ground state energy was computed as a power series in the coupling $\alpha$, defined in \eqref{a-tilde} 
(although $\xi=0$ in this case). At leading order in the $1/N$ expansion, one obtains for
$\CE_{(0)}(\alpha)$ the same result that we presented in \eqref{E0-alpha}. At next-to-leading order in $1/N$ one finds
\be
\label{se1-ba}
\CE_{(1)}^\text{susy} (\alpha)= -\frac{21 \zeta (3)}{32} \alpha^4 + \frac{35 \zeta
   (3)}{32}\alpha^5- \left(\frac{735 \zeta (3)}{512}+\frac{4185 \zeta
   (5)}{2048}\right) \alpha^6+ \CO\left(\alpha^7\right), 
\ee
which is available up to order $\alpha^{42}$ in \cite{mr-ren}. Interestingly, this expansion is almost identical to the bosonic result \eqref{e1-ba}, but  
it only keeps its transcendental part.
 
 Our goal in this section 
 is to test the result \eqref{se1-ba} against a perturbative calculation. Like before, it is more convenient to use the linearized version of the 
model, which is obtained by introducing three auxiliary fields: one scalar field $X$ to impose the constraint \eqref{cons}, 
a Majorana fermion $\lambda$ to impose the constraint \eqref{2cons}, and a Hubbard--Stratonovich scalar field $\tau$ to integrate out the quartic fermionic term in the Lagrangian \eqref{susyL}. The resulting Euclidean Lagrangian, which includes the coupling to the chemical potential $h$ for both $\boldsymbol S$ and $\boldsymbol\Psi$ fields, is given by\footnote{In Euclidean signature, the gamma matrices satisfy the anti-commutator relation $\{\gamma^\mu,\gamma^\nu\} = 2 \eta_E^{\mu \nu} \mathbb{I}$, where $\eta_E^{\mu \nu}$ is the Euclidean metric. After a Wick rotation of the action, the Euclidean gamma matrices $\gamma_E$ that enter in the Euclidean Lagrangian \eqref{susyLh} are connected to the Minkowski matrices $\gamma_M$ by $\gamma^0 = \gamma^0_M$ and  $\gamma^j = \ri \gamma^j_M$, for $j \neq 0$.}
\begin{multline}
\CL_h^\text{susy} = \frac{1}{2g_0^2}\bigg\{\partial_\mu \bS\cdot \partial^\mu\bS+2\ri h (S_1\partial_0 S_2-S_2 \partial_0 S_1) + h^2(S_3^2+\dots+S_N^2-1)+X(\bS^2-1)\\
+ \widebar{\bPsi} \cdot \slashed{\partial} \bPsi + 2\widebar{\lambda} \bS \cdot\bPsi+ \tau^2+\tau \widebar{\bPsi}\cdot\bPsi- \ri h\left(\widebar{\Psi}_1\gamma_0\Psi_2-\widebar{\Psi}_2\gamma_0\Psi_1\right)\bigg\}.
\label{susyLh}
\end{multline}

\begin{figure}
\centering
\includegraphics{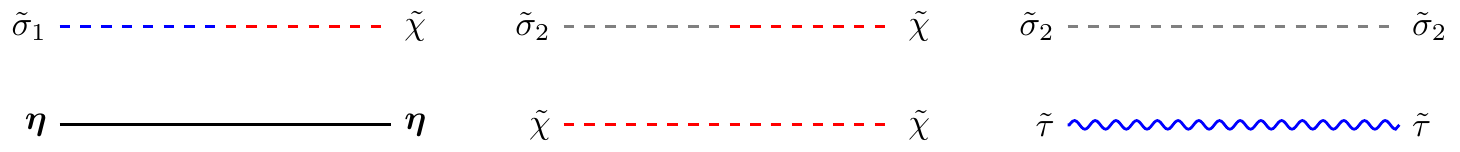}
\caption{Propagators for bosonic fields in the supersymmetric non-linear sigma model.}
\label{bosonprops}
\end{figure}
\begin{figure}
\centering
\includegraphics{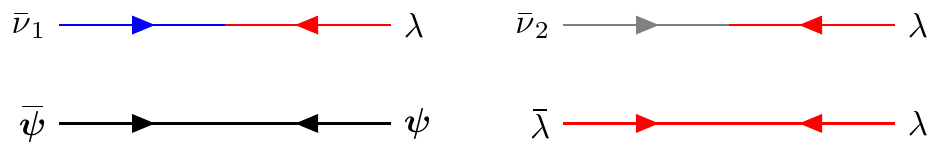}
\caption{Propagators for fermionic fields in the supersymmetric non-linear sigma model.}
\label{fermionprops}
\end{figure}

\begin{figure}
\centering
\includegraphics{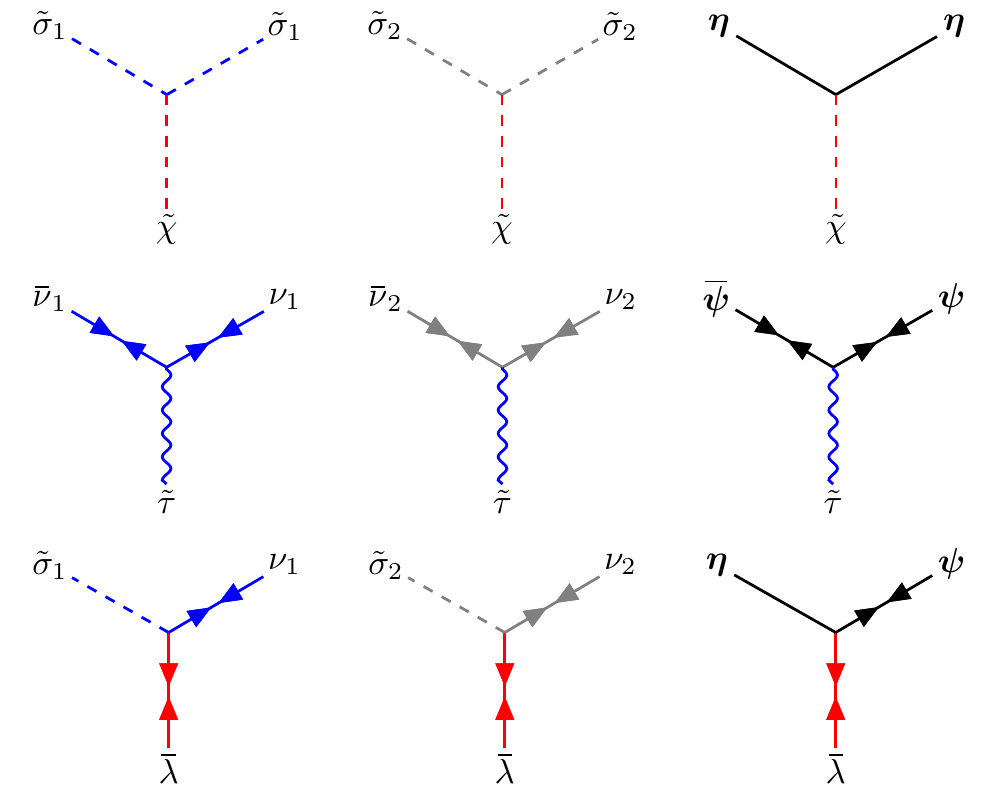}
\caption{Interaction terms in the supersymmetric non-linear sigma model.}
\label{s-vertices}
\end{figure}

We now expand the Lagrangian around the following classical vacuum, 
\begin{equation}
\ba 
\boldsymbol S (x) &= \Big(\sigma,0,\dots,0\Big) + \sqrt{2 \pi \Delta \lambda_0 } \Big( \tilde{\sigma}_1(x),\tilde\sigma_2(x),\eta_1(x),\dots,\eta_{N-2}(x)\Big),\\
\boldsymbol\Psi(x) &= \Big(0,\dots,0\Big) + \sqrt{2 \pi \Delta \lambda_0} \Big( \nu_1(x),\nu_2(x), \psi_1(x),\dots,\psi_{N-2}(x)\Big),\\
X(x) &= \chi + \sqrt{2 \pi \Delta \lambda_0} \tilde\chi(x), \qquad
\tau(x) = \tau + \sqrt{2 \pi \Delta \lambda_0}\tilde{\tau}(x), \qquad
\lambda(x) =  0 + \sqrt{2 \pi \Delta \lambda_0} \lambda(x).
\ea
\end{equation}
Up to linear terms in the fields, one finds that the Lagrangian can be written as
\begin{equation}
\mathcal{L}_h^\text{susy} = {1 \over 2 \pi  \Delta \lambda_0} \CL^\text{susy}_{\text{tree}} + \CL^\text{susy}_{\text{G}}+\sqrt{2 \pi \Delta \lambda_0 } \CL^\text{susy}_{\text{int}}. 
\end{equation}
The tree-level Lagrangian is given by
\be
\CL^\text{susy}_{\text{tree}} = \frac{\chi}{2}(\sigma^2-1)+\frac{\tau^2}{2}-\frac{h^2}{2}. 
\label{susyLtree}
\ee
The quadratic terms are given by
\begin{multline}
\CL^\text{susy}_{\text{G}} =
\frac{1}{2} \boldsymbol\eta\cdot(-\partial^2+\chi+h^2)\boldsymbol\eta 
+ \frac{1}{2}\tilde{\tau}^2 +
\frac{1}{2} \begin{bmatrix}
\tilde{\sigma}_1\\ \tilde{\sigma}_2\\ \tilde{\chi}
\end{bmatrix}^T
\begin{pmatrix}
- \partial^2+\chi & 2 \ri h \partial_0 & \sigma\\
 -2 \ri h \partial_0 & - \partial^2+\chi & 0\\
\sigma & 0 & 0
\end{pmatrix}
\begin{bmatrix}
\tilde{\sigma}_1\\ \tilde{\sigma}_2\\ \tilde{\chi}
\end{bmatrix}
\\
+\frac{1}{2}\widebar\bpsi \cdot(\slashed{\partial}+\tau)\bpsi + \frac{1}{2}
\begin{bmatrix}
\widebar{\nu}_1\\ \widebar{\nu}_2\\ \widebar{\lambda}
\end{bmatrix}^T
\begin{pmatrix}
\slashed{\partial} + \tau &  -\ri h \gamma_0 & \sigma\\
  \ri h \gamma_0 & \slashed{\partial}+\tau & 0\\
\sigma & 0 & 0
\end{pmatrix}
\begin{bmatrix}
\nu_1\\ \nu_2\\ \lambda
\end{bmatrix},
\label{susyLkin}
\end{multline}
where $\bpsi = (\psi_1,\dots,\psi_{N-2})$. Finally, the interaction terms are given by
\be
\CL^\text{susy}_{\text{int}} =
\frac{1}{2}\tilde{\chi}\boldsymbol\eta\cdot \boldsymbol\eta
+ \frac{1}{2}\tilde{\chi} \Big( \tilde{\sigma}_1^2 + \tilde{\sigma}_2^2 \Big)
+ \widebar{\lambda} \Big( \tilde{\sigma}_1\nu_1 + \tilde{\sigma}_2\nu_2 + \boldsymbol\eta\cdot\bpsi\Big)
+ \frac{1}{2}\tilde{\tau} \Big( \widebar{\nu}_1\nu_1 + \widebar{\nu}_2\nu_2 + \widebar\bpsi \cdot\bpsi \Big).
\label{susyLint}
\ee
From \eqref{susyLkin} we can compute the propagators in momentum space. The propagators of the boson fields $\boldsymbol\eta$, $\tilde\sigma_1$, $\tilde\sigma_2$ and $\tilde\chi$ were already obtained in \eqref{nlsmLkin}. There is, however, an additional boson $\tilde\tau$ with propagator
\begin{equation}
D_{\tilde{\tau}\tilde{\tau}} = 1.
\end{equation}
The fermion propagators are
\begin{equation}
\ba 
S_{\widebar{\nu}_1\nu_1} &= S_{\widebar{\nu}_2\nu_1} = S_{\widebar{\nu}_1\nu_2} = 0, & S_{\widebar{\nu}_2\nu_2} &=  \frac{1}{\ri\slashed{k}+\tau},\\
S_{\widebar{\nu}_1\lambda} &= S_{\widebar{\lambda}\nu_1} =  \frac{1}{\sigma}, & S_{\widebar{\nu}_2 \lambda} &=\frac{\ri h}{\sigma}\gamma_0\frac{1}{\ri\slashed{k}+\tau}, & S_{\widebar{\lambda} \nu_2} &=-\frac{\ri h}{\sigma}\frac{1}{\ri\slashed{k}+\tau}\gamma_0,\\
S_{\widebar{\lambda}\lambda} &= -\frac{1}{\sigma^2}\left[\ri\slashed{k}+\tau - \gamma_0 \frac{ h^2}{\ri\slashed{k}+\tau}\gamma_0 \right], & S_{\widebar{\psi}_i\psi_j}&= \delta_{ij}  \frac{1}{\ri\slashed{k}+\tau}, & i,j&=1, \dots, N-2. 
\ea
\label{eq_fermprops}
\end{equation}
The propagators of the bosonic and the fermionic fields are represented diagrammatically as in \figref{bosonprops} and \figref{fermionprops}. The interaction terms are represented by the vertices in \figref{s-vertices}.

We can now calculate the effective potential in an expansion in 
powers of $\Delta$, as in \eqref{v-delta}. The leading order term comes from the tree-level Lagrangian together with the one-loop contributions of the $\eta$ bosons and $\psi$ fermions, for which there are $1/\Delta = N-2$ of each:
\begin{equation}
V_{(0)}(\sigma,\chi,\tau) = \frac{1}{4\pi \lambda_0}\left[\chi(\sigma^2-1)+\tau^2-h^2\right]+\frac{1}{2}\int\frac{\rd^d k}{(2\pi)^d} \log\left(\frac{k^2+h^2+\chi}{k^2+\tau^2}\right).
\end{equation}
By using dimensional regularization, we obtain
\begin{equation}
V_{(0)}(\sigma,\chi,\tau;h)= \frac{1}{4\pi\lambda_0}\left[\chi(\sigma^2-1)+\tau^2-h^2\right]+\frac{(h^2+\chi)^{d/2}}{(4\pi)^{d/2}}\frac{1}{d}\Gamma\left(\frac{\epsilon}{2}\right)-\frac{(\tau^2)^{d/2}}{(4\pi)^{d/2}}\frac{1}{d}\Gamma\left(\frac{\epsilon}{2}\right).
\end{equation}
The extremization procedure for $\chi$ and $\sigma$ is identical to the bosonic case. As for $\tau$, much like $\chi$, there is the non-perturbative choice $\tau\neq 0$ and the trivial one $\tau=0$. We choose the latter to connect our result with perturbation theory. After renormalizing the coupling like in the bosonic case, we find
\begin{equation}
\delta\CF_{(0)}^\text{susy}(h) = - \frac{h^2}{4\pi}\left\{\frac{1}{\lambda}+\log\left(\frac{h}{\mu}\right)-\frac{1}{2}\right\}.
\end{equation}

\subsection{Ring diagrams}

There are three types of ring diagrams that contribute to the effective potential at next-to-leading order in $\Delta$: ring diagrams 
with bosonic $\eta$ bubbles, ring diagrams with fermionic $\psi$ bubbles, and ring diagrams with mixed 
$\psi$-$\eta$ bubbles. The first type of ring diagrams 
are the same ones of the bosonic sigma model, so we do not have to compute them again. Since the propagators for $\psi$ and $\tilde\tau$ 
are both $h$ independent, fermionic ring diagrams do not contribute to $\delta\CF(h)$ after subtraction at $h=0$. Thus we only need to consider mixed ring diagrams.

The $\psi$-$\eta$ bubbles are connected by $\lambda$ lines. Then the contribution of mixed ring diagrams to the effective potential is
\begin{equation}
\sum_{m\geq 1} \frac{1}{2m}\int \frac{\rd^d k}{(2\pi)^d} \Tr\left[ \left( 2\pi\lambda_0 S_{\widebar{\lambda}\lambda}(k)\Pi^\mu_{\psi\eta}(k,h^2+\chi,\tau)\gamma_\mu\right)^m \right].
\end{equation}
A minus sign arises from the single fermion loop that runs across the entire diagram. This sign gets canceled by an additional sign that appears in the computation of the effective potential. The polarization loop is in this case 
\begin{equation}
\Pi^\mu_{\psi\eta}(k,M^2,\tau) = \int \frac{\rd^d q}{(2\pi)^d} \frac{1}{q^2+ M^2} \frac{-\ri (k^\mu+ q^\mu) + \tau}{(k+q)^2 + \tau^2}.
\end{equation}
This integral can be computed with standard Feynman techniques and, after evaluation at the vacuum $\chi_{(0)} = \tau_{(0)} = 0$, we obtain
\begin{equation}
\begin{aligned}
\Pi^\mu_{\psi\eta}(k,h^2,0) &= -\ri k^\mu \Pi_{\psi\eta}(k^2,h^2),\\
\Pi_{\psi\eta}(k^2,h^2) &=
\frac{\Gamma(1+\frac{\epsilon}{2})}{(4\pi)^{d/2}}\int_0^1 \rd y \frac{y^{-\epsilon/2}}{\big[h^2 + (1-y)k^2\big]^{1+\epsilon/2}}\\
&= h^{-2-\epsilon} \frac{\Gamma(1+\frac{\epsilon}{2})}{(4\pi)^{d/2}} \frac{(1+x)^{-1-\epsilon/2}}{1-\epsilon/2} {}_2F_1\left(1+\frac{\epsilon}{2},1-\frac{\epsilon}{2};2-\frac{\epsilon}{2};\frac{x}{x+1} \right).
\end{aligned}
\label{polarization-loop-susy}
\end{equation}
Again, as in the bosonic case, we have expressed the polarization loop in terms of a hypergeometric function and the variable $x=k^2/h^2$. 
The free energy at next-to-leading order in $\Delta$ from bosonic plus mixed ring diagrams can now be written as
%\
%\begin{equation}
%\Pi_{\psi\eta}(k^2) 
%= \frac{\Gamma(1+\frac{\epsilon}{2})}{(4\pi)^{d/2}}\int_0^1 \rd y \frac{y^{-\epsilon/2}}{\big[h^2 + (1-y)k^2\big]^{1+\epsilon/2}},
%\end{equation} 
%and
\begin{equation}
\delta\CF_{(1)}^\text{susy}(h) = -\frac{h^2\nu^{-\epsilon}}{4\pi}\left[ \frac{2\pi\nu^\epsilon}{h^2} \sum_{m\geq 1}\frac{(-1)^m}{m} \left( \frac{2\pi\lambda_0}{\sigma_{(0)}^2} \right)^m (\CI_m - \mathfrak{I}_m) \right].
\label{eq_V_psi_eta}
\end{equation}
The integrals $\CI_m$, corresponding to bosonic diagrams, were already defined in \eqref{eq_V1_Im}, and we 
have a set of new integrals from the mixed diagrams given by
\begin{equation}
\mathfrak{I}_m = \int \frac{\rd^d k}{(2\pi)^d} \left[\Pi_{\psi\eta}(k^2,h^2)\right]^m \Tr\left[ \left( k^2 + \frac{h^2}{k^2}\gamma_0\slashed{k}\gamma_0\slashed{k} \right)^m \right].
\label{Im-susy}
\end{equation}
The front factor in \eqref{eq_V_psi_eta} has been extracted for better comparison with the 
non-supersymmetric result of \eqref{renV1}. As we already mentioned, one important difference in 
the present model is that we do not need a renormalization constant 
%contributing at next-to-leading order 
to cancel the divergences of the ring diagrams. Instead, there is a total cancellation of divergences 
between bosonic and mixed diagrams. We will see this explicitly in section \ref{sec_susysing}, thus proving 
that the beta function at subleading order is $\beta_{(1)}(\lambda) = 0$.

\begin{figure}
\centering
\begin{subfigure}[b]{.32\linewidth}
\includegraphics{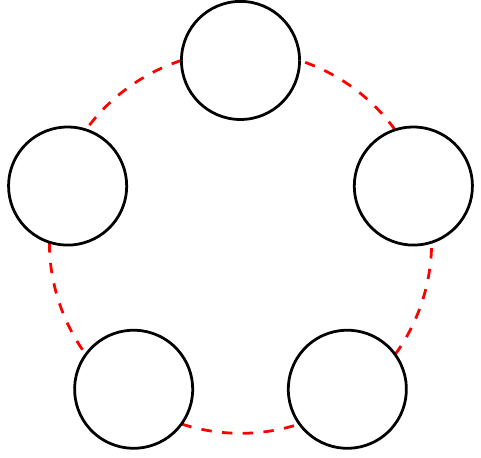}
\caption{Bosonic\label{ring-a}}
\end{subfigure}
\begin{subfigure}[b]{.32\linewidth}
\includegraphics{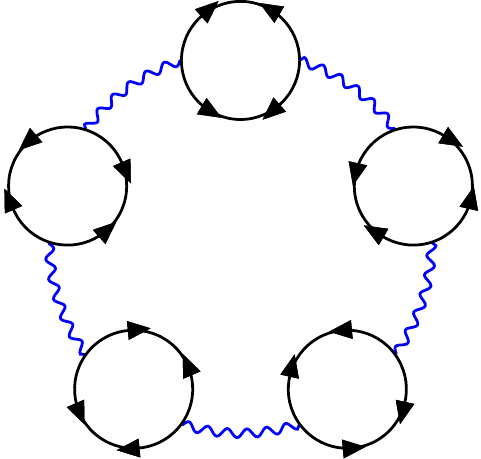}
\caption{Fermionic\label{ring-b}}
\end{subfigure}
\begin{subfigure}[b]{.32\linewidth}
\includegraphics{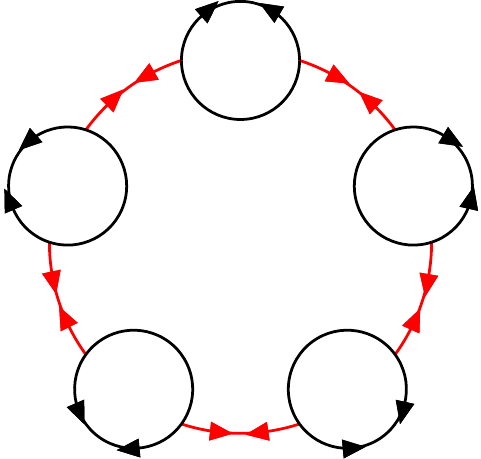}
\caption{Mixed\label{ring-c}}
\end{subfigure}
\caption{The three types of ring diagrams that appear at order $\Delta^0$. 
\subref{ring-a} Ring diagram from the bosonic non-linear sigma model. \subref{ring-b} Ring diagram with pure fermion bubbles. This type of diagrams do not depend on $h$. \subref{ring-c} Ring diagrams with mixed 
boson-fermion bubbles, which we compute in this section.}
\label{rings}
\end{figure}

At this stage, it is convenient to compute the trace in \eqref{Im-susy}. Expanding with the binomial theorem, we obtain
\begin{equation}
\Tr\left[ \left( k^2 + \frac{h^2}{k^2}\gamma_0\slashed{k}\gamma_0\slashed{k} \right)^m \right]
= \sum_{\ell=0}^m\binom{m}{\ell} (h^2)^{\ell} k^{2m-4\ell} \Tr\left[(\gamma_0\slashed{k})^{2\ell}\right].
\label{trace-computation}
\end{equation}
We can calculate the trace recursively in arbitrary dimension $d$:
\begin{equation}
\Tr\left[(\gamma_0\slashed{k})^{2\ell}\right] = 2^{\lfloor\frac{d}{2}\rfloor}\left[\frac{1}{2}\left(k_0+\ri \sqrt{k^2-k_0^2}\right)^{2\ell}+\frac{1}{2}\left(k_0-\ri \sqrt{k^2-k_0^2}\right)^{2\ell}\right].
\end{equation}
As is the standard procedure, we take the dimension of the spinor representation to be a fixed integer 
$2^{\lfloor\frac{d}{2}\rfloor}=2$. In $d$-dimensional spherical coordinates we can always pick 
$k_0 = k \cos \theta_1$. Then we obtain
\begin{equation}
\Tr\left[(\gamma_0\slashed{k})^{2\ell}\right]
%= \left[\left(k_0+\ri \sqrt{k^2-k_0^2}\right)^{2\ell}+\left(k_0-\ri \sqrt{k^2-k_0^2}\right)^{2\ell}\right]
= 2 k^{2\ell} \cos(2 \ell \theta_1),
\end{equation}
and the momentum angular integral can now be computed term by term in \eqref{trace-computation}, yielding
\begin{equation}
\mathfrak{I}_m = h^{2-\epsilon(m+1)} \frac{S_d}{(2\pi)^d} \left[ \frac{\Gamma(1+\frac{\epsilon}{2})}{(4\pi)^{d/2}} \right]^m \sum_{\ell=0}^m \binom{m}{\ell} \frac{(-1)^\ell \Gamma(\frac{d}{2})^2}{\Gamma(\frac{d}{2}-\ell)\Gamma(\frac{d}{2}+\ell)}\mathfrak{I}_{m,\ell},
\label{Im-sum}
\end{equation}
where we singled out the integrals
\begin{equation}
\mathfrak{I}_{m,\ell} = \frac{1}{(1-\frac{\epsilon}{2})^m}\int_0^1 \rd z \frac{z^{m-\ell-\epsilon/2}}{(1-z)^{2-\ell-(m+1)\epsilon/2}} \left[ {}_2F_1\left( 1 + \frac{\epsilon}{2}, 1 - \frac{\epsilon}{2}; 2 - \frac{\epsilon}{2}; z  \right) \right]^m.
\label{eq_Iml_susy}
\end{equation}
For $\ell\geq 2$ the integrals are finite at $\epsilon=0$. Since the factor $1/\Gamma(d/2-\ell)$ vanishes in the sum of \eqref{Im-sum} for $\epsilon=0$, none of the terms with $\ell \ge 2$ will contribute to the free energy.

\subsection{Cancellation of divergences}
\label{sec_susysing}
The goal in this section is to check that the subleading free energy \eqref{eq_V_psi_eta} in the 
supersymmetric model is already finite without the need of renormalization. For that, we follow similar 
techniques to those in section \ref{ring-diagrams} and \ref{beta-singular-part}. That is, we want to isolate the singular part of the 
integrals $\mathfrak{I}_{m,\ell}$, as we did in \eqref{bml-exp}--\eqref{bmlk}. We start by expressing the 
hypergeometric function in \eqref{eq_Iml_susy} as a sum of two terms, by using fractional linear transformations:
\begin{multline}
\label{flinear}
{}_2F_1\left( 1 + \frac{\epsilon}{2}, 1 - \frac{\epsilon}{2}; 2 - \frac{\epsilon}{2}; z  \right) =
-\frac{2-\epsilon}{\epsilon} \frac{\Gamma(1-\frac{\epsilon}{2})^2}{\Gamma(1-\epsilon)} z^{-1+\epsilon/2}\\
\times\left[ 1 - \frac{\Gamma(1-\epsilon)}{\Gamma(1-\frac{\epsilon}{2})^2} (1-z)^{-\epsilon/2} {}_2F_1\left( \frac{\epsilon}{2}, -\frac{\epsilon}{2}; 1 - \frac{\epsilon}{2}; 1-z  \right) \right].
\end{multline}
We now go back to \eqref{eq_Iml_susy}, plug in (\ref{flinear}) and change the variable of integration from $z$ to $1-z$. We get,
\begin{align}
\mathfrak{I}_{m,\ell} &= \left(-\frac{2}{\epsilon} \frac{\Gamma(1-\frac{\epsilon}{2})^2}{\Gamma(1-\epsilon)} \right)^m \mathfrak{B}_{m,\ell},\\
\mathfrak{B}_{m,\ell} &= \int_0^1 \rd z \frac{(1-z)^{-\ell+(m-1)\epsilon/2}}{z^{2-\ell-(m+1)\epsilon/2}}\left[ 1 - \frac{\Gamma(1-\epsilon)}{\Gamma(1-\frac{\epsilon}{2})^2} z^{-\epsilon/2} {}_2F_1\left( \frac{\epsilon}{2}, -\frac{\epsilon}{2}; 1 - \frac{\epsilon}{2}; z  \right) \right]^m. \label{bml-susy}
\end{align}
By expanding the square bracket with the binomial theorem and integrating term by term with the Euler beta function, we obtain
\begin{align}
\mathfrak{B}_{m,\ell} &= \sum_{k\ge 0} \mathfrak{B}_{m,\ell,k},\\
\mathfrak{B}_{m,\ell,k} &=\sum_{s=0}^m \binom{m}{s} \left(-\frac{\Gamma(1-\epsilon)}{\Gamma(1-\frac{\epsilon}{2})^2}\right)^s
\frac{\Gamma\left(-\ell + (m-1)\frac{\epsilon}{2} + 1\right)\Gamma\left(k + \ell + (m-s+1)\frac{\epsilon}{2} - 1\right)}{\Gamma\left(k + (2m-s)\frac{\epsilon}{2}\right)} d_k^{(s)}.
\label{eq_Bml_beta_susy}
\end{align}
The coefficients $d_k^{(s)}$ are defined by the Taylor expansion
\begin{equation}
\left[ {}_2F_1\left( \frac{\epsilon}{2}, -\frac{\epsilon}{2}; 1 - \frac{\epsilon}{2}; z  \right) \right]^s = \sum_{k\ge 0} d_k^{(s)} z^k.
\end{equation}
The Laurent expansion of $\mathfrak{B}_{m,\ell}$ up to order $\epsilon^m$ is obtained 
by summing only the terms $k=0$, $1$ for $\ell=0$ and $k=0$ for $\ell=1$. This follows 
from a computation similar to the one in appendix \ref{computation_bhat}. We then find,
\begin{equation}
\begin{aligned}
\mathfrak{B}_{m,0} &= \mathfrak{B}_{m,0,0} + \mathfrak{B}_{m,0,1} + \CO\left(\epsilon^{m+1}\right),\\
\mathfrak{B}_{m,1} &= \mathfrak{B}_{m,1,0} + \CO\left(\epsilon^{m+1}\right).
\end{aligned}
\label{bml-epsilon-susy}
\end{equation}
%
%This result is slightly different to the non-supersymmetric case (compare \eqref{bml-epsilon-susy} to \eqref{bml-epsilon}), and it greatly simplifies the computation of mixed ring diagrams, as it means that the terms in \eqref{bml-epsilon-susy} already incorporate the singular part plus the finite part of $\mathfrak{I}_{m,\ell}$.
This result is slightly different to the bosonic case (compare \eqref{bml-epsilon-susy} to \eqref{bml-epsilon}), and it greatly simplifies the computation of mixed ring diagrams, because  the terms in \eqref{bml-epsilon-susy} already incorporate the singular part plus the finite part of $\mathfrak{I}_{m,\ell}$.

Equation \eqref{eq_Bml_beta_susy} gives a divergent result for $m=1$, $\ell=1$ (even for arbitrary $\epsilon$). In this special case we have to compute the integral \eqref{eq_Iml_susy} explicitly. Instead of using the hypergeometric representation, we use the integral representation of \eqref{polarization-loop-susy} and commute the $x$ and $y$ integrals. Then we can compute both integrals analytically in terms of the Euler beta function $B(x,y)$. We obtain, 
\begin{equation}
\mathfrak{I}_{1,1} = B\left(1-\frac{\epsilon}{2},\epsilon \right)B\left(1-\frac{\epsilon}{2},\frac{\epsilon}{2} \right) = \frac{2}{\epsilon ^2}+\frac{\pi ^2}{4}+\CO\left(\epsilon ^1\right).
\end{equation}

%We can also easily compute the integral $m=1$, $\ell=0$ with the same trick. In this case,
%\begin{equation}
%\mathfrak{I}_{1,0} = B\left(2-\frac{\epsilon}{2},-1 + \frac{\epsilon}{2} \right) B\left(1-\frac{\epsilon}{2},-1 + \frac{\epsilon}{2} \right) = 0\,.
%\end{equation}

We are now ready to compute the divergent part arising from mixed ring diagrams in \eqref{eq_V_psi_eta}. We start by considering the sum
\begin{equation}
\mathfrak{S}_m = \mathfrak{B}_{m,0,0} + \mathfrak{B}_{m,0,1} + \frac{m\epsilon}{2-\epsilon}\mathfrak{B}_{m,1,0},
\label{SB_susy}
\end{equation}
which can be written in the form
\begin{equation}
\mathfrak{S}_m = \Gamma\left(\frac{m-1}{2}\epsilon\right)\sum_{s=0}^m \binom{m}{s} \frac{(-1)^s}{m-s+1} \mathfrak{f}_m(s\epsilon,\epsilon),
\end{equation}
with
\begin{equation}
\begin{aligned}
\mathfrak{f}_m(y,x) &=
\frac{\Gamma\left(\frac{m+1}{2}x-\frac{y}{2}+1\right)}{\Gamma\left( mx - \frac{y}{2}\right)}\left( \frac{2(m-1)}{(m+1)x- y - 2}
+ \frac{yx(m-1)}{(x-2)(2mx-y)} + \frac{2m}{2-x} \right) \re^{y\mathfrak{h}(x)},\\
\mathfrak{h}(x) &= \frac{1}{x}\log\left( \frac{\Gamma(1-x)}{\Gamma(1-\frac{x}{2})^2} \right). 
\end{aligned}
\end{equation}
We can now perform the sum in $s$, as we did in (\ref{divpart}), and we obtain the result
\begin{equation}
\mathfrak{S}_m = \frac{2(-1)^m}{m+1} \left(\frac{\Gamma(1-\epsilon)}{\Gamma(1-\frac{\epsilon}{2})^2}\right)^{m+1} \frac{\epsilon-1}{\epsilon-2} - \frac{2(-1)^m}{(m+1)(m-1)}\bigg[\frac{\rd^{m+1} \mathfrak{f}_m(y,0)}{\rd y^{m+1}} \bigg]_{y=0}\epsilon^m + \CO\left(\epsilon^{m+1}\right).
\label{divpart-susy}
\end{equation}
The first term in the r.h.s. is equal to $\mathfrak{S}_m$ up to order $\epsilon^{m-1}$, so that term is sufficient to compute the singular part of the integrals $\mathfrak{I}_m$. The second term, contributing to order $\epsilon^m$, will be needed for the computation of the finite part of $\mathfrak{I}_m$.

At this point, it becomes a simple exercise to verify that the singular parts of $\CI_m$ and $\mathfrak{I}_m$ cancel each other in \eqref{eq_V_psi_eta}. We first observe that
\begin{equation}
\mathfrak{S}_m = 2\cdot 2^{-\epsilon(m+1)} \CS_m + \CO\left(\epsilon^m\right).
\label{eq_ffrakS}
\end{equation}
If we now rewrite the sum of the terms $\ell=0$, $1$ appearing in \eqref{Im} and \eqref{Im-sum} using $\CS_m$ and $\mathfrak{S}_m$, respectively, one can verify from the observation above that
\begin{equation}
\CI_m = \mathfrak{I}_m + \CO\left(\epsilon^0\right).
\end{equation}
This proves that the Laurent expansion of \eqref{eq_V_psi_eta} starts at the constant term. In other words, there is a complete cancellation of divergent terms between bosonic and mixed ring diagrams. In addition, this shows that $\beta_{(1)}(\lambda)=0$ from a direct computation in perturbation theory.

\subsection{Finite part of the free energy and comparison with the Bethe ansatz}

We can also organize the finite parts of the mixed ring diagrams contributing to the free energy in a manner similar to what we did in \eqref{def_WXYZ} for bosonic ring diagrams. We note, however, that there are a few simplifications in this case. 
The analogue of $Z$, corresponding to the sum of the integrals $\mathfrak{I}_{m,\ell}$ for $\ell\ge 2$, is equal to 0. The analogue of $X$, corresponding to the contribution from the terms we missed in the sum \eqref{SB_susy}, is also 0, as is easily inferred from the result \eqref{bml-epsilon-susy}. Moreover, the analogue of $W$ is present in the mixed ring diagrams, but it exactly cancels with 
that of the bosonic diagrams (this can be deduced from the derivation that led to \eqref{eq_ffrakS}).

In short, we only need to consider the analogue of $Y$, which arises from the second term in the r.h.s. of \eqref{divpart-susy}. 
After simplifying $\mathfrak{f}_m(y,0)$ we find
\begin{equation}
 \mathfrak{f}_m(y,0) 
= \frac{y}{2}\left( \frac{m-1}{y/2+1} - m \right),
\end{equation}
and the computation of the $(m+1)$-th derivative follows naturally from the geometric series. Then, 
the contribution to the free energy is simply
\begin{equation}
\mathfrak{Y}\left(\lambda;\frac{h}{\mu}\right)= \frac{1}{2\pi}\sum_{m\geq 1} \frac{(m-1)!}{2^m}\left(\frac{\lambda}{1+\lambda \log\left(h/\mu\right)}\right)^m.
\label{eq_fracY}
\end{equation}
Putting all our results together we obtain
\begin{equation}
\delta\CF_{(1)}^{\text{susy}}(h) =  -\frac{h^2}{2}\bigg\{X\left(\lambda;\frac{h}{\mu}\right)+Y\left(\lambda;\frac{h}{\mu}\right)+Z\left(\lambda;\frac{h}{\mu}\right)+\mathfrak{Y}\left(\lambda;\frac{h}{\mu}\right)\bigg\},
\end{equation}
%where $X$, $Y$ and $Z$ are defined in \ef{def_WXYZ}.
Thus, the ground state energy is
\begin{equation}
\CE_{(1)}^{\text{susy}}(\alpha) = -2\pi \alpha^2 \big[ X\left(\alpha;1\right)+Y\left(\alpha;1\right)+Z\left(\alpha;1\right)+\mathfrak{Y}\left(\alpha;1\right)\big].
\label{gse_delta_susy}
\end{equation}
Here, $\alpha$ is defined by the equation \eqref{a-tilde} with $\beta_0$ given in (\ref{susy-beta}) and $\xi=0$.  We find perfect agreement between \eqref{gse_delta_susy} and the coefficients obtained from the Bethe ansatz, which were calculated up to order $\alpha^{42}$ in \cite{mr-ren}.

As we previously observed in \eqref{se1-ba}, the perturbative expansion of $\CE_{(1)}^{\text{susy}}(\alpha)$ does not contain any rational term (at least to the available order). In our perturbative computation, we verified that $\mathfrak{Y}(\alpha;1)$ cancels the rational part of $X\left(\alpha;1\right)+Y\left(\alpha;1\right)$, so that the coefficients are linear combinations of Riemann zeta functions evaluated at odd arguments.

As we did in the bosonic case, we can obtain from these results the location of the renormalon singularities and their associated trans-series. The mixed diagrams \eqref{eq_fracY} contribute only to the IR renormalon singularity at $\zeta=2$, and we find
 \begin{multline}
  \Im\CE_{(1)}^{\text{susy}}(\alpha) =
\frac{\pi}{2}\bigg[ \alpha^2 \re^{-2/\alpha} +\left(4 - \alpha ^2\right) \re^{2/\alpha} +\left(\frac{108}{\alpha ^2}+\frac{72}{\alpha }\right)\re^{6/\alpha}\\
 +\left(\frac{50000}{9 \alpha ^4}+\frac{20000}{3 \alpha ^3}+\frac{6500}{3 \alpha ^2}+\frac{200}{\alpha }\right) \re^{10/\alpha} + \CO\left(\re^{14/\alpha}\right)\bigg],
  \label{disc_E_SUSY}
 \end{multline}
 where only the sign of the IR term differs from the bosonic case. The asymptotic behaviour extracted from this discontinuity 
 matches the coefficients with the expected precision. Let us note that, although divergences cancel between bosonic and mixed ring diagrams, the IR renormalon at $\zeta=2$ does not cancel. This is in contrast to the cancelation of leading IR renormalons that occurs in some supersymmetric theories according to \cite{shifman-susy, dsu}. 

\section{Conclusions}
\label{conclude-sec}
%we had fun. we saw a renormalon. 

The Bethe ansatz calculation of the free energy in two-dimensional integrable models is one of the most interesting exact results 
in quantum field theory. It makes it possible to understand quantitatively many important aspects of asymptotically free theories, 
like dynamical mass generation and the presence of renormalons and condensates. For this reason, it is important to test the predictions of the 
Bethe ansatz against more conventional methods in quantum field theory. This has been done in the past, in particular in the case of the 
Gross--Neveu model \cite{fnw1,fnw2}. However, a direct test against perturbation theory was limited until very recently by the difficulty of extracting 
perturbative series from the Bethe ansatz. The results of \cite{volin, volin-thesis,mr-ren} have opened the possibility of such a test, and this 
has been the first goal of this paper. We have performed a calculation of the free energy at next-to-leading order in the $1/N$ expansion and at all loops, and we have verified the predictions of the Bethe ansatz up to very high order in the coupling constant, both 
in the non-linear sigma model and its 
supersymmetric extension. The second goal of this paper has been to use this analytic, all-orders result to obtain detailed information about renormalon singularities and their associated trans-series, at leading order in the $1/N$ expansion. 

There are many problems open by our investigation and by closely related efforts. First of all, one could extend the tests presented here to other 
integrable field theories. In \cite{mr-ren}, explicit results for the perturbative series of the free energy have been obtained for the Gross--Neveu model and the principal chiral field with different choices of chemical potentials. There are also results for integrable, non-relativistic models, like the 
Gaudin--Yang model \cite{mr-long}. All these perturbative expansions, obtained from the Bethe ansatz, could be tested by conventional techniques, and these tests would provide additional insights. It should be mentioned that, in the case of the principal chiral field, which is a matrix model, a direct perturbative calculation at higher loops is more challenging than for the vector models that we studied in this paper. 
Another interesting problem is to extend our results to higher orders in the $1/N$ expansion, even for the models considered here. 

In our calculation of the free energy we did an expansion around the trivial large $N$ vacuum, 
in order to make contact with perturbation theory. However, 
it is well known that both the non-linear sigma model and its supersymmetric version have a non-trivial large $N$ saddle point, which leads to a 
non-perturbative mass gap. In fact, previous analysis of renormalons in the non-linear 
sigma model have been based on 
expansions around the non-trivial large $N$ saddle point \cite{david1, david2, itep}. A preliminary calculation of the free energy 
considered in this paper around this non-trivial vacuum seems to lead to a purely non-perturbative result, 
at next-to-leading order in the $1/N$ expansion. 
It would be very interesting to understand more precisely the relationship between the 
expansions around these two very different vacua, and the r\^ole of renormalons in each of them. Note that similar issues are raised in the 
two-dimensional linear $O(N)$ model. There, the ground state energy can be calculated in the trivial large $N$ vacuum, similar to what we did in this paper \cite{mr-new}, but it can be also calculated in the non-trivial large $N$ vacuum \cite{serone}. 

The exact Bethe ansatz solution contains in principle all the information in the problem, both perturbative and 
non-perturbative. One of the goals of the 
resurgence program is to ``semiclassically decode" this exact answer, by writing it as a Borel--\'Ecalle resummation of a trans-series. 
Substantial evidence that this can be done was obtained recently in \cite{abbh1,abbh2}, in the case of the $O(4)$ sigma model, by an impressive 
calculation. Additional evidence has been given in the context of integrable many-body systems, in \cite{mr-long,mr-hub}. 
The first step in the semiclassical decoding is to obtain the explicit form of the trans-series. So far, in all the problems solved by the Bethe ansatz, this has been done by looking at the large order behavior of the perturbative sector. A more challenging problem is to extract the trans-series directly from the integral equation defining the free energy. This would require an extension of Volin's method by explicitly including the exponentially suppressed corrections. 

Another important open question is to provide a physical interpretation of the 
trans-series that we have obtained. It was argued in \cite{mr-ren} 
that the IR renormalon at $\zeta=2$ in the non-linear sigma model might be explained 
by the condensate of the operator $\CO= \partial_\mu \boldsymbol{S}\cdot \partial^\mu \boldsymbol{S}$. Is 
it possible to devise some sort of perturbation theory in the background of the condensate which allows us to reproduce analytically the trans-series? Such a generalized perturbation theory, akin to the one used in the calculation of OPEs \cite{sum-rules}, would provide one of the 
missing ingredients in our understanding of quantum field theory.

\section*{Acknowledgements}
We would like to thank Lorenzo di Pietro, Marco Serone and Giacomo Sverbeglieri for useful 
discussions. This work has been supported in part by the Fonds National Suisse, 
subsidy 200020-175539, by the NCCR 51NF40-182902 ``The Mathematics of Physics'' (SwissMAP), 
and by the ERC-SyG project 
``Recursive and Exact New Quantum Theory" (ReNewQuantum), which 
received funding from the European Research Council (ERC) under the European 
Union's Horizon 2020 research and innovation program, 
grant agreement No. 810573.

\appendix

\section{Integral representation of the series \texorpdfstring{$\widehat{\CB}_{m,\ell}$}{B ml}}
\label{computation_bhat}

In this appendix we will derive the integral representation of $\widehat{\CB}_{m,\ell}$ that we presented in \eqref{habmell}. Because the result has a Laurent expansion starting with $\epsilon^m$, this will also prove that the integrals (\ref{bml}) for $\ell=0$, $1$ satisfy
\begin{equation}
\begin{aligned}
\CB_{m,0} &= \CB_{m,0,0} + \CB_{m,0,1} + \CO\left(\epsilon^m\right),\\
\CB_{m,1} &= \CB_{m,1,0} + \CO\left(\epsilon^m\right).
\end{aligned}
\label{bml-epsilon}
\end{equation}
where the coefficients $\CB_{m, \ell, k}$ are given in (\ref{bmlk}). In particular, this means that the 
singular part of the integrals $\CI_m$ can be extracted from the combination in \eqref{SB}. 

We start from \eqref{bml} and expand the binomial. We then subtract the terms $k=0$, $1$ (for $\ell=0$) or $k=0$ (for $\ell=1$) to obtain the following expression for the sums in \eqref{bhat}:
\be
\widehat\CB_{m,\ell} = \int_0^1 \rd z \frac{(1-z)^{m/2-\ell-\epsilon/2}}{z^{2-\ell-(m+1)\epsilon/2}} \sum_{s=0}^m \binom{m}{s}(-1)^s g_\ell(s\epsilon,\epsilon;z) \label{bhat-integral}. 
\ee
In this equation, 
\be
g_\ell(y,x;z) = z^{-y/2} \re^{y \mathfrak{g}(x)} \left[ \re^{y \Phi(x, z)} - 1 +  \delta_{\ell,0}{z y \over 2(2-x)}  \right],
\ee
where $\mathfrak{g}$ is given in \eqref{doublef}, and $\Phi(x, z)$ is defined by 
\be
\Phi(x, z)= {1\over x} \log \left[ {}_2 F_1 \left(-{x \over 2}, {1\over 2}; 1-{x \over2}; z \right) \right]= 
\log \left[ {1 + {\sqrt{1-z}} \over 2} \right] + \CO(x). 
\ee
We now Taylor expand $g_\ell(y,x;z)$ in its first argument
\begin{equation}
g_\ell(y,x;z) = \sum_{r\ge 0}\frac{y^r}{r!}  \left[ \frac{\rd^r g_\ell(y,x;z)}{\rd y^r}\right]_{y=0}.
\end{equation}
The binomial identity
\begin{equation}
\sum_{s=0}^m \binom{m}{s} (-1)^s s^r =
\begin{cases}
0, & 0 \le r < m,\\
(-1)^m m!, & r=m,
\end{cases}
\end{equation}
allows us to perform the sum over $s$ in \eqref{bhat-integral} for each term of the Taylor expansion. Every term with $r < m$ vanishes after the binomial sum, and we obtain
\begin{equation}
\label{bhat-inter}
\widehat\CB_{m,\ell} = \epsilon^m (-1)^m \int_0^1 \rd z \frac{(1-z)^{m/2-\ell-\epsilon/2}}{z^{2-\ell-(m+1)\epsilon/2}} \left[ \frac{\rd^m g_\ell(y,\epsilon;z)}{\rd y^m}\right]_{y=0} + \CO\left(\epsilon^{m+1}\right).
\end{equation}
The $\CO\left(\epsilon^{m+1}\right)$ contains further terms of the Taylor expansion that we have ignored. Because the above 
integral is convergent at $\epsilon=0$, we can set $\epsilon=0$ in the integrand of (\ref{bhat-inter}) (further corrections will be of order $\epsilon^{m+1}$). This completes the proof of \eqref{bml-epsilon} and yields the result in \eqref{habmell}, 
where $g_\ell(y;z) \equiv g_\ell(y,0;z)$.
 
\section{Analytic computation of the integrals 
%contributing to \texorpdfstring{$X$}{X} and \texorpdfstring{$Z$}{Z}}
}
\label{finite-ints}
%\texorpdfstring{\boldmath $\CI_{m,\ell}$}{I ml} with \texorpdfstring{\boldmath $\ell\ge 2$}{l>=2}}

In this appendix we derive and present analytic results for the integrals $\CI_{m,\ell}$ and $\widehat\CB_{m,\ell}$ appearing in \eqref{def_WXYZ}.  

Let us start with the integrals $\widehat\CB_{m,\ell}$, defined in \eqref{habmell}. One way to compute them is to obtain a closed expression for the generating functional
\begin{equation}
g_{m,\ell}(y) \equiv \int_0^1 \mathrm{d}z \frac{(1-z)^{m/2-\ell}}{z^{2-\ell}} g_\ell(y,z), \qquad \ell=0, 1,
\end{equation}
and then extract the desired integral by taking the $m$-th Taylor coefficient of $g_{m,\ell}(y)$ around $y=0$.

First we fix $\ell = 1$, in which case we have the closed expression
\begin{equation}
\ba
g_{m,1}(y) &= 2\frac{ \Gamma (m+1) \Gamma (-\frac{y}{2})\, {}_2F_1\left(m,1-\frac{y}{2};m-\frac{y}{2};-1\right)}{m \Gamma (m-\frac{y}{2})}-\frac{2^{y+1} \Gamma (\frac{m}{2}+1) \Gamma (-\frac{y}{2})}{m \Gamma (\frac{m}{2}-\frac{y}{2})}.
\ea
\label{f1_hypg}
\end{equation}
This expression is not suited for a Taylor expansion around $y=0$ (because of the hypergeometric function), but it can be put in a more appropriate form by observing that
\begin{equation}
_2F_1\left(m,1-\frac{y}{2};m-\frac{y}{2};-1\right) = \frac{1}{2} \, _2F_1\left(-\frac{y}{2},m-1;-\frac{y}{2}-(m-1)+(2 m-1);-1\right).
\end{equation}
We then use Theorem 3 in \cite{2F1thms} and expand the singular gamma functions around their poles, which leads to the expression 
\begin{multline}
_2F_1\left(a,j;a-j+m;-1\right) = 2^{-a}\sqrt{\pi }\frac{  \Gamma (a-j+m) }{\Gamma (j) \Gamma(m-j)}\\
\times \sum _{k=0}^{m-1} (-1)^{k+j}\binom{m-1}{k}\frac{2^{-k} \Gamma (a+k)}{\Gamma (a) \Gamma (\frac{a+k+1}{2}) } \frac{ \psi ^{(0)}(\frac{a+k}{2}-j+1)}{ \Gamma(\frac{a+k}{2}-j+1)} ,
\label{2f1intj}
\end{multline}
where $\psi^{(0)}$ is the polygamma function. 
After some additional massaging, we obtain the following alternative expression for \eqref{f1_hypg}, in terms of Pochhammer symbols:
\begin{multline}
g_{m,1}(y)= 
\frac{(-1)^m}{2 \Gamma (m-1)} \sum _{k=0}^{2 m-2}  
(-1)^{k+1} \binom{2 m-2}{k} \left(\frac{k}{2}-m-\frac{y}{4}+2\right)_{m-2} \psi ^{(0)}\left(\frac{k}{2}-m-\frac{y}{4}+2\right)\\
-\frac{2^y \Gamma(\frac{m}{2}) \Gamma(-\frac{y}{2})}{\Gamma(\frac{m-y}{2})}.
\label{gm1}
\end{multline}
For $\ell=0$ and $m\geq 2$, a similar calculation leads to
%
%\begin{equation}
%\ba
%f_{m,0}(y) &=
%\frac{1}{8} \bigg( 
%\frac{ 2^{\frac{y}{2}+2} \sqrt{\pi }}{\Gamma (m-1)}
%\sum _{k=0}^{2 m} \binom{2 m}{k} \frac{2^{-k} (-1)^{k+m-1} \Gamma \left(k-\frac{y}{2}-1\right) \psi ^{(0)}\left(\frac{1}{2} \left(k-\frac{y}{2}-1\right)-m+2\right)}{\Gamma \left(\frac{1}{2} \left(k-\frac{y}{2}\right)\right) \Gamma \left(\frac{1}{2} \left(k-\frac{y}{2}-1\right)-m+2\right)}\\
%&-\frac{2^y (4 m+(y-2) y) \Gamma \left(\frac{m}{2}+1\right) \Gamma \left(-\frac{y}{2}-1\right)}{\Gamma \left(\frac{m}{2}-\frac{y}{2}+1\right)}\bigg).
%\ea
%\end{equation}
\begin{multline}
g_{m,0}(y) =
-\frac{1}{8\Gamma (m-1)}\sum _{k=0}^{2 m} (-1)^{k}\binom{2 m}{k}\left(-\frac{k}{2}+\frac{y}{4}+\frac{3}{2}\right)_{m-2}\psi ^{(0)}\left(\frac{k}{2}-m-\frac{y}{4}+\frac{3}{2}\right) \\
-\frac{2^y (4 m+(y-2) y) \Gamma (\frac{m}{2}+1) \Gamma (-\frac{y}{2}-1)}{8\Gamma (\frac{m}{2}-\frac{y}{2}+1)}.
\label{gm0}
\end{multline}
The expressions \eqref{gm1} and \eqref{gm0} are now ready to be Taylor expanded around $y=0$. From the Taylor coefficients we can then extract analytic expressions for the integrals in \eqref{habmell}.

Let us now consider the integrals $\CI_{m, \ell}$, defined in \eqref{Iml}, with $2 \le \ell \le m$. We first focus on the particular case $\ell=m$: $\CI_{m,m}$. Knowing that the integrals converge without the need of regularization, 
we can set $\epsilon=0$, and the hypergeometric function in the integrand becomes
\begin{equation}
{}_2F_1\left(1, \frac{1}{2}; \frac{3}{2}; z \right) = \frac{\tanh^{-1}(\sqrt{z})}{\sqrt{z}}.
\end{equation}
Then, after the change of variable $x=\tanh^{-1}(\sqrt{z})$, the integral \eqref{Iml} at $\epsilon=0$ can be written as
\begin{equation}
\CI_{m,m} = \frac{1}{2} \int_0^\infty \frac{x^m \re^{-(m-1)x/2}}{(1-\re^{-x})^{m-1}}\mathrm{d}x.
\label{eq_Imm}
\end{equation}
By expanding the denominator and integrating term by term, we obtain a sum of Hurwitz zeta functions:
\be
\CI_{m,m} = \frac{m!}{2}\sum_{i=0}^{m-2} b_i \zeta\left(m-i+1,\frac{m-1}{2}\right), 
\label{eq_Imm_semi_result}
\end{equation}
where
\be
b_i= \frac{1}{(m-2)!}\sum _{j=i}^{m-2}  \binom{j}{i} \left(\frac{m-3}{2}\right)^{j-i} S_{m-2}^{(j)}
\end{equation}
and $S_m^{(i)}$ are the Stirling numbers of the first kind. The Hurwitz zeta function appears with integer values in the first argument and both integer and half-integer values in the second argument. In these cases, the zeta function reduces to
\begin{align}
\zeta(n,k) &= \zeta(n) - H_{k-1}^{(n)}, \qquad k\in\mathbb{Z},\\
\zeta(n,k) &= (2^n-1)\zeta(n) - 2^n - \sum_{q=0}^{k-3/2} \frac{1}{(q+\frac{1}{2})^n}, \qquad k + \frac{1}{2}\in \mathbb{Z},
\label{HurZ}
\end{align}
where $H_{k}^{(n)}$ is the $k$-th harmonic number of order $n$, and $\zeta(z)$ is the Riemann zeta function. This completes the analytic computation of $\CI_{m,m}$. However, our result \eqref{eq_Imm_semi_result} can be written in a compact form by using the conventions of umbral calculus:
\begin{equation}
\CI_{m,m} =\frac{m(m-1)}{2} \zeta ^{m+1} \prod _{i=0}^{m-3} \left(\frac{1}{\zeta}-\frac{m-3}{2}+i\right), \qquad \zeta^n \mapsto \zeta\left(n,\frac{m-1}{2}\right).
\label{eq_Imm_closed}
\end{equation}
In these conventions, we first expand the product as a finite polynomial in an abstract variable $\zeta$, and then we replace each power of $\zeta$ by the Hurwitz zeta function, according to the mapping in \eqref{eq_Imm_closed}.

By using similar tricks, we can also compute $\CI_{m,\ell}$, with $\ell=2,\dots,m-1$, in closed form. However, from the computational point of view, it is more efficient to instead consider the following recursion formula:
\begin{equation}
\CI_{m,\ell} = \frac{2}{\ell-1} \Big[ (m-2\ell)\,\CI_{m,\ell+1} + m\, \CI_{m-1,\ell} \Big], \qquad m \ge3, \quad \ell \ge 2,
\label{recIml} 
\end{equation}
which can be obtained by integrating \eqref{Iml} by parts. Using this recursion and the values $\CI_{m,m}$ for $m \ge 2$ as initial data, one obtains all convergent integrals $\CI_{m, \ell}$.  

One can use this derivation to obtain results on the transcendentality of the integrals. By observing that the function
\begin{equation}
\sum_{i=0}^{m-2} b_i x^i = -\frac{\sin \left(\pi x+(m-3)\frac{\pi}{2}\right)}{\pi(m-2)!}
\Gamma \left(\frac{m}{2}-x-\frac{1}{2}\right) \Gamma \left(\frac{m}{2}+x-\frac{1}{2}\right)
\end{equation} 
is odd for odd $m$ and even for even $m$, one shows that the $b_i$ vanish for $m-i+1$ even. Thus, only Hurwitz zeta functions $\zeta(n,\frac{m-1}{2})$ where $n$ is odd show up in \eqref{eq_Imm_semi_result}. Furthermore, the rational parts in \eqref{HurZ} are a finite sum of rational numbers, for both $m$ odd and even. By exchanging these sums with the sum over $i$ in \eqref{eq_Imm_semi_result}, one can easily show that the total rational part vanishes. Thus $\CI_{m,m}$ is a linear combination of the values $\zeta(2k+1)$, with $k\in \mathbb{N}$ and $1\leq k \leq m/2$. Thanks to the recursion relation \eqref{recIml}, the same statement trivially applies to all $\CI_{m,\ell}$ with $2\leq \ell \leq m$.

\section{Asymptotic expansions of the discontinuities}
\label{app-disc}

To obtain the discontinuity of $X(\lambda;1)$, as presented in \eqref{disc_X}, we first have to 
express the residues in \eqref{discX_form} in terms of the position of the poles $z_1$ and $z_2$. This can be easily done 
with L'Hôpital's rule. Then it suffices to compute $z_1$ and $z_2$ perturbatively in powers of $\re^{2/\lambda}$. These perturbative 
solutions can be conveniently obtained by defining variables $v$ and $\xi$ such that
\be
z=1-(1-2\xi v)^2, \qquad \xi=\re^{2/\lambda},
\label{def_v}
\ee
and define $v_i$ such that $z_i=1-(1-2\xi v_i)^2$. The variable $\xi$ in (\ref{def_v}) should not be confused 
with the parameter introduced in (\ref{xi-def}). From \eqref{def_z12} we obtain
\begin{equation}
v_1 = \left(\xi  v_1^2+\re ^{\frac{2 \xi v_1   }{1-2 \xi  v_1}\log \xi}\right), \qquad
v_2 = \left(\xi +\re ^{-\frac{2 \xi  v_2 }{1-2 \xi  v_2}\log \xi}\right)^{-1}.
\label{def_v12}
\end{equation}
We now write the perturbative solution as
\begin{equation}
v_i = 1+ v_{i}^{(1)} \xi  + v_{i}^{(2)} \xi^2 + \dots,
\end{equation}
If we plug this ansatz in \eqref{def_v12} we can calculate the coefficients $v_i^{(j)}$ recursively, and we find
\begin{equation}
\ba 
v_1 &= 1+\xi  (2 \log\xi+1)+\xi ^2 \left(6 \log ^2\xi+10 \log\xi+2\right) +\CO\left(\xi ^3\right),\\
v_2 &= 1+\xi  (2 \log\xi-1)+\xi ^2 \left(6 \log ^2\xi -2 \log \xi +1\right)+\CO\left(\xi ^3\right).
\ea
\label{res_v12}
\end{equation}

To compute the discontinuity of $Y(\lambda;1)$ presented in \eqref{disc_Y}, one distinguishes 
the case $\lambda>0$ from the case $\lambda<0$. We recall that the exponential integral $\text{E}_1(x)$ can be written as an 
entire function plus $-\log(x)$. %Thus the discontinuity of $\text{E}_1$ is $-2\pi \ri$ when its argument is negative. 
In the case of $\lambda>0$, we find that the argument of $\text{E}_1$ is negative for all $z\in(0,1)$, and we obtain simply
\begin{equation}
\disc Y(\lambda>0; 1) = -\frac{1}{4\pi}\int_{0}^{1}\frac{\rd z}{z(1-z)} \frac{z \re^{-\frac{2}{\lambda  \sqrt{1-z}}}}{\lambda  \sqrt{1-z}}(-2\pi \ri) = \frac{\ri\re^{-2/\lambda}}{2}.
\end{equation}
When $\lambda<0$ the calculation is similar to the one for $X(\lambda;1)$. We have a logarithmic discontinuity when $z\in(0,z_1)$, and a pole at $z=z_1$. In total we find
\begin{equation}
\ba
\disc Y(\lambda<0;1) &=
-2\pi\ri \Bigg[\frac{1}{4\pi}\int_{0}^{z_1}\frac{\rd z}{z(1-z)} \frac{z \re^{-\frac{2}{\lambda  \sqrt{1-z}}}}{\lambda  \sqrt{1-z}}+\frac{1}{4\pi}\Res\left(\CY(\lambda\sqrt{1-z},z),z=z_1\right)\Bigg]\\
&= -\frac{\ri}{2}\left[\re^{-2/\lambda}-\re^{-\frac{2}{\lambda \sqrt{1-z_1}}} + \frac{z_1 + 2}{\lambda z_1 (1-z_1)} \left[\frac{\rd}{\rd z_1}\sqrt{1-z_1} \log \left(\frac{\sqrt{z_1}}{2}\right)\right]^{-1} \right] ,
\ea
\label{discY-computation}
\end{equation}
where we have to plug the $z_1$ solution in \eqref{def_v} and \eqref{res_v12}, leading to \eqref{disc_Y}.
 
For the discontinuity of $Z(\lambda;1)$, it is convenient to change variables as
\begin{equation}
z = (1-2 \xi v)^2.
\end{equation}
From \eqref{def_z3} we find that
\begin{equation}
z_3 = (1- 2 \xi v_2)^2,
\end{equation}
where $v_2$ is the same as in \eqref{def_v12} and \eqref{res_v12}. The asymptotic approximation to the residue of the pole in \eqref{Zlambda} now follows in the same way as for the pole of $Y(\lambda;1)$. Meanwhile, from \eqref{def_z4} we obtain a much more simple equation for $v_4$, 
\begin{equation}
v_4 = \re^{-2\xi \log \xi v_4}(1-\xi v_4),
\end{equation}
which is the same equation found in the closely related calculations of \cite{mr-hub,mr-roads}. Thus, we can take the same solution
\begin{equation}
v_4 = 1 - \xi  (2 \log (\xi )+1)+ \xi ^2 \left(6 \log^2\xi+6 \log\xi+1\right) + \CO\left(\xi ^3\right).
\label{res_v4}
\end{equation}
We are left with computing the discontinuity of the logarithm term in \eqref{Zlambda}. This discontinuity, given by \eqref{disc}, comes from the branch cut of the square root when $\CZ(z,\lambda)<0$. Following the strategy of \cite{mr-hub,mr-roads}, we define the functions 
\begin{equation}
\ba
A_2(v) &= \begin{multlined}[t]
2  (v_2 - v)\xi \log\xi + 2\log\xi+\log\left(\frac{v_2-\xi v v_2}{v-\xi v v_2}\right)\\
+4\xi\left[ (v_2- \xi v^2_2) \log \left(\frac{1-\xi  v_2}{\xi  v_2}\right)- (v- \xi v^2) \log \left(\frac{1-\xi  v}{\xi  v}\right)\right],
\end{multlined}\\
A_4(v) &= 2 (v_4 - v) \xi \log\xi  + \log\left(\frac{v_4-\xi v v_4}{v-\xi v v_4}\right).\\
\ea
\end{equation}
By construction we have that $A_i(v_i)=0$ for $i=2$, $4$. In terms of these functions, the discontinuity in \eqref{disc} can be written as
\begin{equation}
\frac{1}{2\pi\xi}\int_{v_4}^{v_2} \rd v 
\frac{1-2 \xi v}{v^2(1-\xi v)^2} \log\left(\frac{\sqrt{A_2(v)}+\sqrt{A_4(v)}}{\sqrt{A_2(v)}-\sqrt{A_4(v)}}\right).
\label{Zdisc_v}
\end{equation}
It is useful to introduce one further variable
\begin{equation}
v= 1+ \xi w,
\label{def_w}
\end{equation}
and $v_i = 1+\xi w_i$.
The advantage of using $A_2$, $A_4$ and $w$ is that we can expand the integrand of (\ref{Zdisc_v}) as a power series in $\xi$, and we find the following structure:
\begin{equation}
\label{serPP}
\sum_{i\geq 0} \left[\CP^{(i)}_l(w,w_2,w_4) \log \left(\frac{\sqrt{w_2-w}+\sqrt{w_4-w}}{\sqrt{w_2-w}-\sqrt{w_4-w} }\right)+  \ri \CP^{(i)}_r(w,w_2,w_4)\frac{\sqrt{\left(w_2-w\right) \left(w-w_4\right)}}{(w_2-w_4)^i}  \right]\xi^i,
\end{equation}  
where $\CP^{(i)}_{l,r}$ are polynomials. For example, up to order $\CO\left(\xi^2\right)$, we find, 
\begin{multline}
\left(1-2 \xi w+O\left(\xi ^2\right)\right) \log \left(\frac{\sqrt{w_2-w}+\sqrt{w_4-w}}{\sqrt{w_2-w}-\sqrt{w_4-w} }\right)\\+  \ri\xi \frac{ \left(8 \log \xi +w_2-w_4+8\right)}{2 \left(w_2-w_4\right)} \sqrt{\left(w_2-w\right) \left(w-w_4\right)} + \CO\left(\xi ^2\right). 
\end{multline}
The coefficients of the $\xi$ series in (\ref{serPP}) are functions of $w$ which are integrable on the interval $[w_4, w_2]$. After integration, 
we replace  $w_2$ and $w_4$ by their respective perturbative solutions through \eqref{res_v12}, \eqref{res_v4} and \eqref{def_w}. Summing up the result of this computation with the residue of the pole at $z=z_3$ leads to \eqref{disc_Z}.

\bibliographystyle{JHEP}

\linespread{0.6}
\bibliography{biblio-nlsm}

\end{document}